\documentclass[numberedappendix]{emulateapj}
\usepackage{amssymb,amsmath,amsthm}
\usepackage{longtable}
\usepackage{amssymb}
\usepackage{amsmath, xspace}
\usepackage{graphics,graphicx} 
\usepackage{rotating}
\usepackage{color}
\usepackage{perpage} 
\renewcommand{\thefootnote}{\arabic{footnote}}
\renewcommand{\thefootnote}{\arabic{footnote}}

\def\xis {{\textit{XIS}}}

\def\apec {{\textit{apec\xspace}\  }}

\def\asca{{\textit{ASCA\xspace}\ }}
\def\rosat{{\textit{ROSAT\xspace}\ }}

\begin{document}


\title{Probing the Outskirts of the Early Stage Galaxy Cluster Merger A1750}


\author{
Esra~Bulbul\altaffilmark{1},
Scott~W.~Randall\altaffilmark{1},
Matthew~Bayliss\altaffilmark{1,2},
Eric~Miller\altaffilmark{3},
Felipe~Andrade-Santos\altaffilmark{1},
Ryan~Johnson\altaffilmark{4},
Mark~Bautz\altaffilmark{3},
Elizabeth~L.~Blanton\altaffilmark{5},
William~R.~Forman\altaffilmark{1},
Christine~Jones\altaffilmark{1},
Rachel~Paterno-Mahler\altaffilmark{5,6},
Stephen~S.~Murray\altaffilmark{1,7},
Craig~L.~Sarazin\altaffilmark{8},
Randall~K.~Smith\altaffilmark{1},
and
Cemile~Ezer\altaffilmark{9}
}
\altaffiltext{1}{Harvard-Smithsonian Center for Astrophysics, 60 Garden Street, Cambridge, MA~02138, USA\\}
\altaffiltext{2}{Department of Physics, Harvard University, 17 Oxford Street, Cambridge, MA 02138\\}
\altaffiltext{3}{Kavli Institute for Astrophysics \& Space Research, Massachusetts Institute of Technology, 77 Massachusetts Ave, Cambridge, MA 02139, USA\\}
\altaffiltext{4}{Department of Physics, Gettysburg College, Gettysburg, PA 17325, USA\\}
\altaffiltext{5}{Astronomy Department and Institute for Astrophysical Research, Boston University, 725 Commonwealth Avenue, Boston, MA 02215, USA\\}
\altaffiltext{6}{Department of Astronomy, University of Michigan, 1085 S. University Avenue, Ann Arbor, MI 48109, USA\\}
\altaffiltext{7}{Department of Physics and Astronomy, Johns Hopkins University, 3400 N. Charles Street, Baltimore, MD 21218, USA\\}
\altaffiltext{8}{Department of Astronomy, University of Virginia, P.O. Box 400325, Charlottesville, VA 22904, USA\\}
\altaffiltext{9}{Department of Physics, Bo\u{g}azi\c{c}i University, Istanbul, Turkey.}

\email{Email:  ebulbul@mit.edu}

\begin{abstract}
We present results from recent {\it Suzaku} and {\it Chandra} X-ray, and {\it MMT} optical observations of the strongly merging ``double cluster"
A1750 out to its virial radius, both along and perpendicular to a
putative large-scale structure filament.  Some previous studies of
individual clusters have found evidence for ICM entropy profiles that
flatten at large cluster radii, as compared with the self-similar
prediction based on purely gravitational models of hierarchical
cluster formation, and gas fractions that rise above the mean cosmic
value. Weakening accretion shocks and the presence of unresolved cool
gas clumps, both of which are expected to correlate with large scale
structure filaments, have been invoked to explain these results.  In
the outskirts of A1750, we find entropy profiles that are consistent
with self-similar expectations, and gas fractions that are consistent
with the mean cosmic value, both along and perpendicular to the
putative large scale filament. Thus, we find no evidence for gas
clumping in the outskirts of A1750, in either direction.
This may indicate that gas clumping is less common in lower temperature ($kT \approx 4$~keV), less massive systems, consistent with some (but not all) previous studies of low mass clusters and groups.
Cluster mass may therefore play a more important role in gas
clumping than dynamical state. Finally, we find evidence for diffuse,
cool ($<$ 1 keV) gas at large cluster radii ($R_{200}$) along the filament, which
is consistent with the expected properties of the denser, hotter phase
of the WHIM.

\end{abstract}
\keywords{X-rays: galaxies: clusters-galaxies: individual (A1750) - X-rays: ICM, WHIM- cosmology: large-scale structure of universe}

\section{Introduction}
\label{sec:intro}

Galaxy cluster mergers are ideal probes of gravitational collapse and the
hierarchical structure formation in the Universe. 
Observations of the evolving cluster mass function provide a sensitive
cosmological test that is both independent of, and complementary to,
other methods (e.g.,  BAO, SN, CMB) \citep{vikhlinin2009}. The use of galaxy clusters as 
cosmological probes relies on the accuracy of scaling relations between 
the total mass and observable quantities. Galaxy cluster mergers will 
disrupt the intracluster gas and cause departures from these scaling
relations \citep[e.g.,][]{randall2002,wik2008}. 
Given that these mass scaling relations are a necessary ingredient for the interpretation of on-going cosmological surveys, a detailed understanding of the intracluster medium
(ICM) gas physics in mergers has become increasingly important.

{\renewcommand{\arraystretch}{1.1}
\begin{table*}[t!]
\begin{center}
\caption{Summary of the {\it Suzaku} and {\it Chandra} X-Ray Pointings  }
\begin{tabular}{llccclcc}
\hline\hline\\
Satellite	& Pointing & ObsID & R.A. & DEC &  Date Obs	& Exposure & PI \\
		&		&		&		&		&	& XIS0/XIS1/XIS3 &\\	
		&		&		&		&		& & (ks)\\
\hline\\
{\it Suzaku}	& North & 806096010 & 13 31 15.53	 & - 01 39 13.3 & 2011 Jul 2 &  74.7/74.7/74.7	& S. Randall\\
{\it Suzaku}	& Center & 806095010 & 13 30 46.63 & - 01 53 14.3 & 2011 Jul 24  & 38.0/38.0/38.0	& S. Randall\\
{\it Suzaku}	& South & 806097010 & 13 30 13.15 & - 02 06 22.7 & 	2011 Jul 9 & 70.2/70.2/70.2	& S. Randall \\
{\it Suzaku}	& Southeast	& 806098010 & 13 31 27.19	 & - 02 04 19.9 & 2011 Jul 6 & 55.9/55.9/56.0	& S. Randall\\
{\it Suzaku}	& Southeast	& 806098020 & 13 31 28.58	 & - 02 02 29.4 & 2011 Dec 23 & 11.3/11.3/11.3	& S. Randall	\\
{\it Chandra}	& North		& 11878	& 13 31 10.83 	 & - 01 43 21.0 & 2010 May 11 & 19.4$^{*}$	& S. Murray	\\
{\it Chandra}	& Center		& 11879	& 13 30 50.30 	& - 01 52 28.0	& 2010 May 9	& 19.7$^{*}$ 	&  S. Murray	\\
{\it Chandra}	& South	& 12914	& 13 30 15.80 	& - 02 02 28.7	& 2011 Mar 16	& 36.8$^{*}$	&  S. Murray	\\
\\
\hline

\multicolumn{5}{l}{%
  \begin{minipage}{7.3cm}%
    \scriptsize $^{*}$ ACIS-I%
  \end{minipage}%
}\\

\hline\hline
\end{tabular}
\label{table:obs}
\end{center}
\end{table*}
}

The properties of the ICM in the cores of merging clusters have been
studied in detail, since the high density and surface brightness of
the gas in this region is well-suited to high angular resolution
observations with {\it Chandra} and {\it XMM-Newton} \citep[see][for a review]{markevitch2007}.
With the launch of the low particle-background {\it Suzaku} mission, it has become possible to probe the low gas density and faint
surface brightness regions at the virial radii of nearby
galaxy clusters \citep[e.g.][]{bautz2009, akamatsu2011, miller2012,urban2014, sato2014}. 
Observational studies at these radii have mostly focused on relatively relaxed, massive, cool-core systems. 
 Due to the limited number of observations, the dynamical evolution of the ICM in
strong merger events out to the viral radius is not clearly
understood.
Strongly merging, bimodal clusters are where we
expect to find the large-scale filaments and accretion
shocks. Comparing results from observations of mergers and relaxed
clusters at the virial radius will provide an important confirmation
of our current picture of large-scale structure formation. The double
clusters identified from \textit{Einstein} observations (A1750, A98,
A115, A3395) are ideal targets for studying the virial radii of
strongly merging clusters \citep{forman1981}. These canonical binary
galaxy clusters have two separated peaks of X-ray emission, and
distortions in their X-ray surface brightness distributions suggest
ongoing merger events \citep[e.g.,][]{pmr2014}. Most of these systems are in fact triple clusters, with all sub-clusters lying roughly along the same line, suggesting the presence of large-scale structure filaments.

A1750 is a triple merger system at a redshift of 0.085, with an
average temperature of 4.5 keV \citep{degrandi2002, neumann2005}. It
contains three main sub-clusters with X-ray centroids: A1750N (J2000, RA: 202.79$^\circ$,
DEC: $-$1.73$^\circ$), A1750C (J2000, RA:
202.71$^\circ$, DEC: $-$1.86$^\circ$), and
A1750S (J2000, RA: 202.54$^\circ$, DEC: $-$2.105$^\circ$). MMT data provided redshifts for the brightest cluster 
galaxies of 0.0836, 0.0878, and 0.0865 (see Section \ref{sec:resultsopt} for details). A1750 was identified as a strongly merging
``double" cluster due to the presence of two bright X-ray subcluster peaks, which are clearly visible in the \textit{Einstein} image \citep{forman1981}. 
 The centers of A1750N and A1750C are separated by 9.7$^\prime$ (930 kpc; see Figure \ref{fig:image}). \asca
and \rosat observations indicate possible shock heated gas with an elevated temperature
of 5.5 keV between these sub-clusters, suggesting that they are in an early stage
merger \citep{donnelly2001}. More recent {\it XMM-Newton} observations confirm
this region of elevated temperature, and also indicate that A1750C may itself be undergoing a
merger \citep{belsole2004}. A1750S was identified with
\rosat observations. Its center is located 17.5$^\prime$ (1.68 Mpc) to the
southwest of A1750C, along the same line connecting A1750C and A1750N, presumably tracing
a large-scale filament. The 0.2 $-$ 10 keV luminosities of the two brighter sub-clusters are 1.3 $\times\, 10^{44}$ ergs s$^{-1}$ for A1750N and 2.2 $\times\, 10^{44}$ ergs s$^{-1}$ for A1750C \citep{belsole2004}. The X-ray luminosity of the fainter, southern sub-cluster A1750S is 6.4 $\times\, 10^{43}$ ergs s$^{-1}$, estimated from \rosat PSPC observations.

Here, we present results from mosaic {\it Suzaku} observations of A1750 out to the virial radius. These new observations, together with archival {\it Chandra} and {\it XMM-Newton}
observations, probe the ICM properties from the subcluster cores out to their viral radii. 
Previous studies of other (non-merging) systems have found entropy profiles 
that flatten at large radii, in contradiction with theoretical predictions, possibly 
due to the presence of unresolved cool gas clumps \citep{urban2014,walker2013}. This behavior shows some 
variation with azimuth, suggesting a connection with large-scale
structure and gas accretion \citep{ichikawa2013,sato2014}. 
We use our observations, which extend both along 
and perpendicular to the putative large-scale structure filament, to look 
for correlations between the ICM properties, the surrounding large-scale environment, and to examine the merger dynamics. This paper is organized as follows: in Section \ref{sec:obs}, we
describe the {\it Suzaku}, {\it Chandra}, and {\it MMT} data used in
our analysis. In Section \ref{sec:analysis}, the analysis of the X-ray and optical observations is described in detail. In Section \ref{sec:syst}, we discuss systematic errors that are relevant to the {\it Suzaku} X-ray measurements at large radii. In Sections \ref{sec:resultsopt},\ref{sec:resultsxray}, and \ref{sec:thermo} we discuss our results and present our conclusions in Section \ref{sec:conc}. Throughout the paper, a standard $\Lambda$CDM cosmology with H$_{0}$ = 70 km s$^{-1}$ Mpc$^{-1}$, $\Omega_{\Lambda}$ = 0.7, and $\Omega_{M}$ = 0.3 is assumed. In this cosmology, 1$^{\prime}$ at the redshift of the cluster corresponds to $\sim$ 96.9 kpc.  Unless otherwise stated, reported errors correspond to 90\% confidence intervals.

\section{Observations and Data Processing}
\label{sec:obs}

\begin{figure}
\centering
\hspace{-4mm}\includegraphics[width=9cm, angle=0]{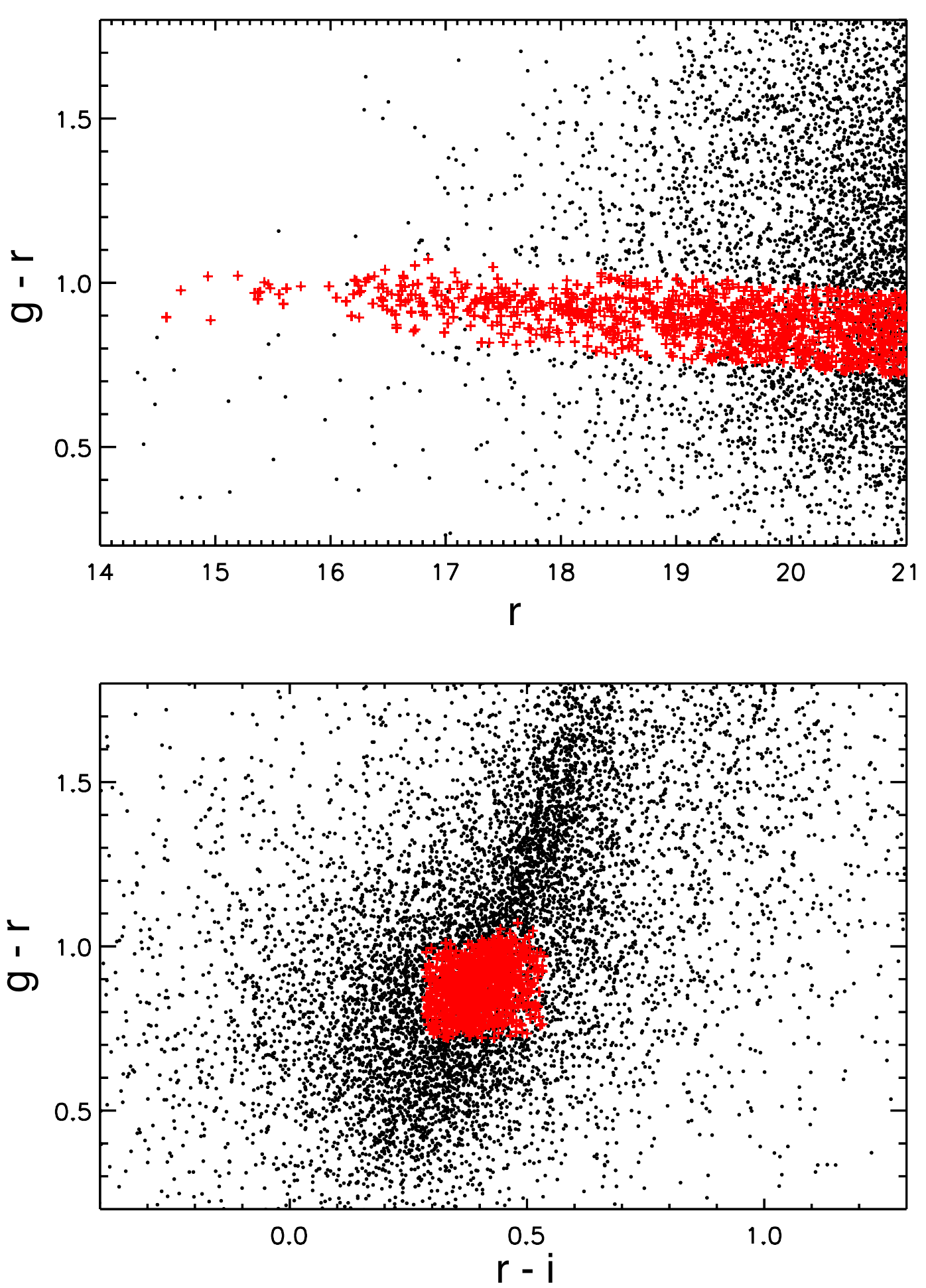}
\vspace{3mm}
\caption{{\bf Upper Panel:} Color-magnitude (g $-$ r vs r) plot of galaxies included in the spectroscopic catalog, with selected passive cluster members plotted in red. {\bf Lower Panel:} Color-color (g $-$ r vs r $-$ i) plot of galaxies included in the spectroscopic catalog, with selected passive cluster members plotted in red.\label{fig:colormag}
}\vspace{2mm}
\end{figure}
%


\subsection{Optical Spectroscopic and Photometric Data}

The majority of the galaxy spectroscopic redshifts used in this analysis are new observations obtained 
using the Hectospec instrument \citep{fabricant2005} at the {\it MMT} Observatory 6.5m telescope 
on Mt. Hopkins, AZ. A single Hectoscpec configuration places up to 300 fibers in a region of 
the sky approximately one degree in diameter. We use data from two such configurations, which 
resulted in 517 individual spectroscopic redshift measurements. 

To supplement our Hectospec spectroscopy, we include data from the literature, when available. 
Specifically, we use 12
spectroscopic redshift measurements from \citet{huchra1995}, 68 from \citet{donnelly2001}, 
seven from \citet{gal2003}, 19 from the 6dF Galaxy Survey \citep{jones2005}, and 200 
from the Sloan Digital Sky Survey \citep[SDSS;][]{ahn2014}. The SDSS
selection includes all objects within a 0.5 degree radius of the centroids of the X-ray 
emission of A1750N, A1750C, and A1750S, and with a spectroscopic redshift falling in the interval 0.03 $< z_{spec} <$ 
0.15, which easily captures the range of recessional velocities of galaxies 
associated with A1750. We then check for duplicate entries across the different 
input redshift catalogs, resulting in 24 removals and a final data set of 799 
spectroscopic redshifts. 

In addition to optical spectroscopy, we also use optical photometry from the SDSS 
catalogs. We perform a query of all objects classified as galaxies within a 0.5 
degree radius of the centroid of the X-ray surface brightness of each subcluster and download all of the 
available optical photometry --- in the \emph{ugriz} bands -- for those 
sources.

\begin{figure*}[]
\centering
\hspace{-1mm}\includegraphics[width=9cm, angle=0]{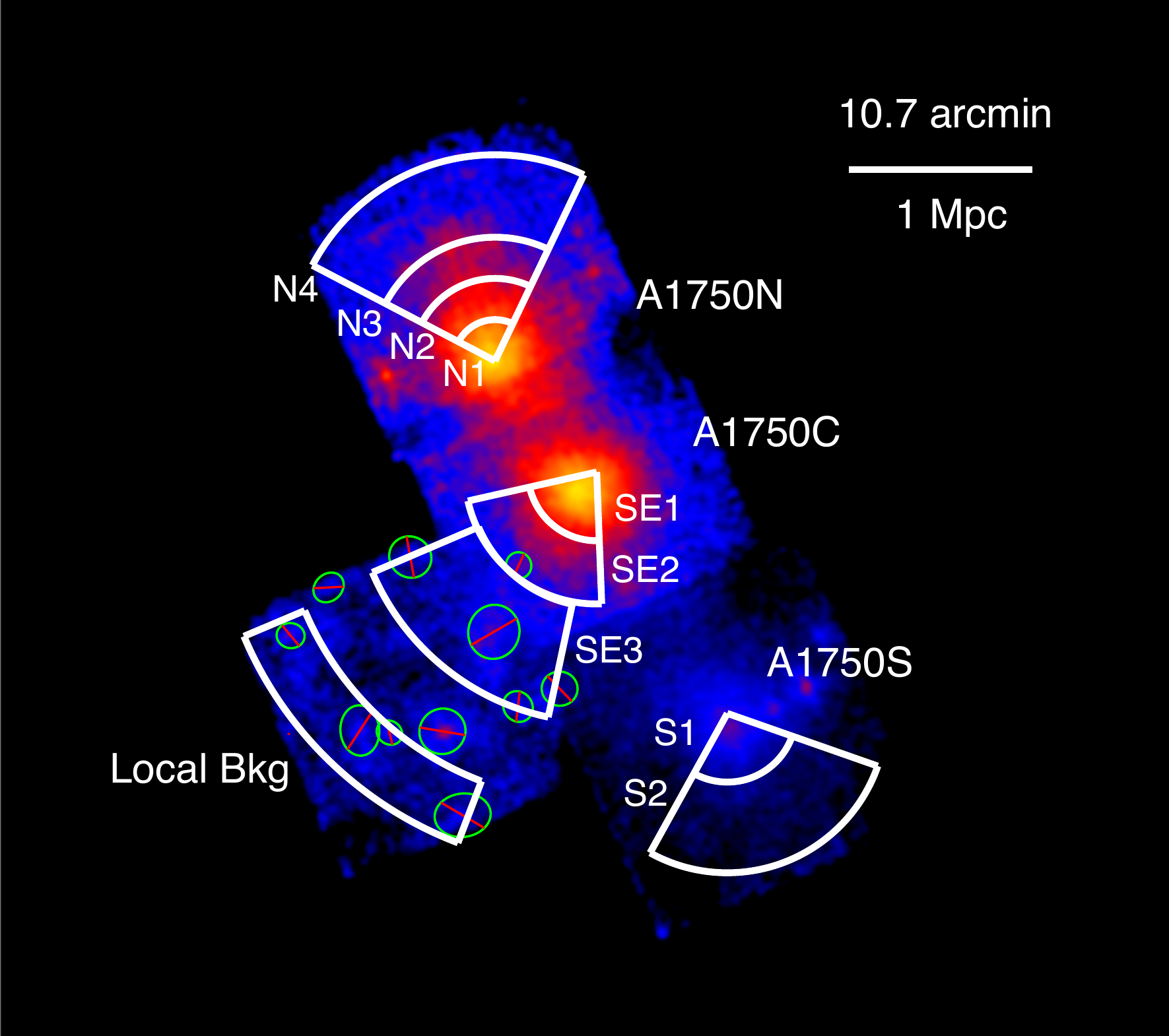}
\hspace{5mm}\includegraphics[width=8.5cm, angle=0]{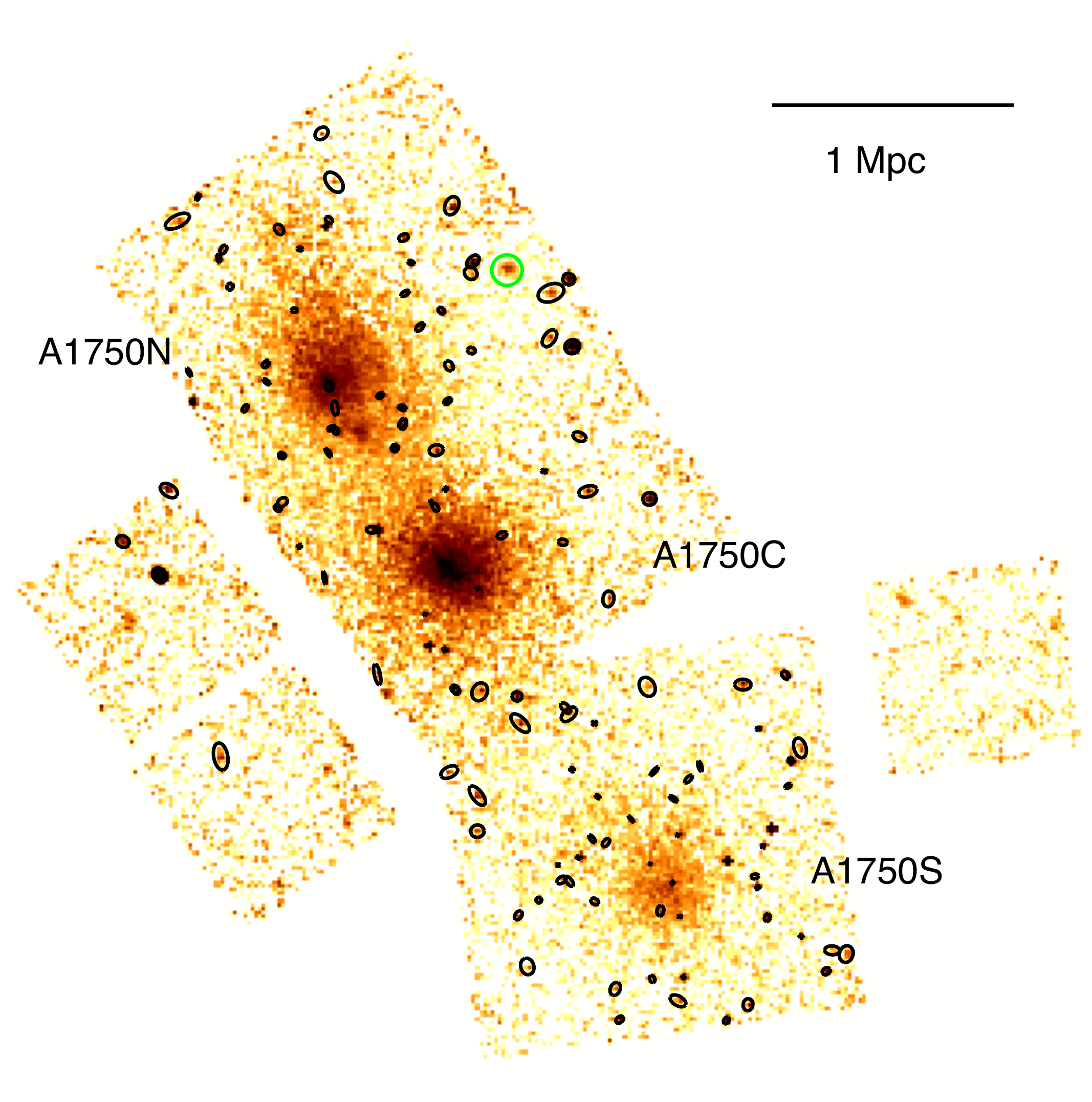}
\vspace{3mm}
\caption{{\bf Left Panel:} Exposure corrected, NXB background subtracted {\it Suzaku} XIS image of the A1750 merger system. The image was extracted in the 0.5 $-$ 7 keV energy range. The spectral extraction regions are shown in white. Point sources in the southeast pointing where we lacked {\it Chandra} observations are shown in green. {\bf Right Panel:} Exposure corrected, background subtracted {\it Chandra} image of A1750. The three {\it Chandra} pointings coinciding with the {\it Suzaku} observations were used to identify the coordinates of the point sources in the {\it Suzaku} field-of-view. The brightest point source, which is in both {\it Suzaku} and {\it Chandra} field-of-view, is shown in green.}
\vspace{4mm}
\label{fig:image}
\end{figure*}
%

\subsection{Suzaku X-ray Observations}
\label{sec:suzaku}

A1750 was observed with {\it Suzaku} with five pointings during July 2011 and
December 2011 (see Table \ref{table:obs}). We process the unfiltered
{\it Suzaku} data with \textit{HEASOFT} version 6.13, and the latest
calibration database CALDB as of  May 2014. The raw event files are
filtered using the FTOOL \textit{aepipeline}. In addition to the
standard filtering performed by \textit{aepipeline}\addtocounter{footnote}{-8},\footnote{http://heasarc.gsfc.nasa.gov/docs/suzaku/processing/criteria\_\\xis.html}
we require an Earth elevation angle $>$ 5$\,^{\circ}$, a geomagnetic
cut-off rigidity of $>$ 6 GV/c, and exclude data collected during
passages through the south Atlantic anomaly as described in
\citet{bautz2009}. The data taken with 3$\times$3 and 5$\times$5
clocking modes are merged and the corners of the chips illuminated by
the Fe-calibration sources are excluded from further analysis. We
carefully examine each light curve after the initial screening to
ensure that the data are free from background flaring events. Due to
the increase in charge injection in data taken with XIS1 after 2011
June 1, the two rows adjacent to the standard charge-injected rows are removed.\footnote{http://heasarc.gsfc.nasa.gov/docs/suzaku/analysis/nxb\_ci6kev.\\html} The region lost due to a putative micrometeorite hit on XIS0 is also excluded from our analysis. The net exposure times of each XIS0, XIS1, and XIS3 pointing after filtering are given in Table \ref{table:obs}. Due to our strict filtering, 30 ks of the total exposure time was lost. The total filtered {\it Suzaku} XIS0/XIS1/XIS3 exposure time is 250.1/250.1/250.2 ks.

\subsection{Chandra X-ray Observations}
\label{sec:chandra}

The {\it Chandra} observations that were used in the analysis are summarized in Table~\ref{table:obs}. For each
observation, the aimpoint was on the front-side illuminated ACIS-I
CCD. All data were reprocessed from the level 1 event files using {\sc CIAO~4.6} and
{\sc CALDB~4.4.7}. CTI and time-dependent gain corrections were
applied. {\sc lc\_clean} was used to check for periods of background
flares.\footnote{\url{http://asc.harvard.edu/contrib/maxim/acisbg/}}
The mean event rate was calculated from a source free region using
time bins within 3$\sigma$ of the overall mean, and bins outside a
factor of 1.2 of this mean were discarded. There were no periods of
strong background flares. To model the background we used the {\sc
CALDB\footnote{\url{http://cxc.harvard.edu/caldb/}}} 
blank sky background files appropriate for this observation, 
normalized to match the 10-12~keV count rate in our observations to
account for variations in the particle background. The total filtered ACIS-I exposure 
time is 75.9 ks.

\section{Analysis }
\label{sec:analysis}

\subsection{Photometric Selection of Cluster Member Galaxies}
\label{sec:photanalysis}

SDSS \emph{ugriz} photometry samples the full optical spectral energy distribution (SED) 
for galaxies in A1750, including the 4000\AA\  break that is located in the 
$g$-band at the redshift of A1750. The 4000\AA\ break is a strong 
feature, characteristic of the passive red sequence galaxies that dominate the galaxy 
populations of evolved galaxy clusters \citep{gladders2000}. We identify candidate 
cluster member galaxies of A1750 using the red sequence in the \emph{gri} bands, 
which span the break. The red sequence selection involves two steps. The initial selection 
is made in color-magnitude space ($g-r$ vs. $r$; Figure~\ref{fig:colormag} top panel) with a manual 
identification of the over-density of galaxies with approximately the same $g-r$ color. We 
then perform a linear fit in color-magnitude space to define the red sequence in 
A1750, and flag all galaxies within $\pm$0.125 in $g-r$ magnitudes as candidate red sequence 
galaxies. The second step occurs in color-color space ($g-r$ vs $r-i$; Figure~\ref{fig:colormag} bottom panel), where we identify an over-density of candidate red sequence galaxies with 
similar $r-i$ colors.

Galaxies that satisfy the initial color-magnitude selection 
while also falling within $\pm$0.125 magnitudes of the mean $r-i$ color of the over 
density in color-color space are flagged as red sequence galaxies. The range of color 
values that we use accounts for both the observed intrinsic scatter in the red sequence 
of massive galaxy clusters 
\citep[$\sim \pm$0.05-0.1 mags;][]{delucia2004,gladders2005,valentinuzzi2011} and the 
typical SDSS photometric uncertainties of $\sim$0.025 magnitudes.

\subsection{Suzaku X-ray  Analysis}
\label{sec:suzanalysis}

We extract an image of A1750 in the 0.5 $-$ 7 keV energy band and
mosaic the pointings in sky coordinates. The non-X-ray background (NXB) 
images are generated using the `night-Earth' data (NTE) using the FTOOL \textit{xisnxbgen} 
\citep{tawa2008}. The NXB images are then subtracted from the mosaicked image prior to exposure correction.

To generate the exposure maps, we first simulate a monochromatic
photon list assuming a 20$^\prime$ uniform extended source for each
observation with the {\it XRT} ray-tracing simulator \textit{xissim}
\citep{ishisaki2007}. These vignetting-corrected photon lists are then
used with \textit{xisexpmapgen} to generate exposure maps of each
pointing, as described in detail in \citet{bautz2009}. Regions with $< 15$\%
of the maximum exposure time are removed. The resulting
exposure maps for each pointing are merged. The particle background subtracted, vignetting-corrected image is shown in Figure \ref{fig:image} (left panel).

\begin{table*}[ht!]
\def\arraystretch{1.5}
\begin{center}
\caption{Contribution of the flux from the adjacent annuli due to PSF spreading and stray light }
\begin{tabular}{lccccccccccc}
\hline\hline\\

 Region & 	N1	&	N2		& N3	& N4	& SE1	& SE2	&SE3	& S1	&	S2		\\
\\
\hline
	\\	
	N1	& 40.88  	&15.64	 	& 1.41	&	0.45 & -	& -		& -	&	-	& -	\\
	N2	& 7.82 	& 56.23		& 15.95	&	1.11& -	& -		& -	&	-	& -	\\
	N3	& 1.18	 & 15.95		& 58.37	&	13.58& -	& -		& -	&	-	& -	\\
	N4	& 0.07	& 0.57		& 8.54	&	55.84& -	& -		& -	& -		&-	\\
	SE1	& -		& -			& -		&	-& 56.64	& 	7.86		& 0.05	& -	& -	\\
	SE2	& -		& -			& -		&	-& 10.81	&  	61.54	& 2.20	& -	& -	\\
	SE3	& -		& -			& -		&	-& 0.45	& 	4.65	& 57.16	& -	& -	\\
	S1 	& -		& -			& -		&	-& -		& -		& -	&	60.78	&	14.35	\\
	S2 	& -		& -			& -		&	-& -		& -		& -	&	9.05	& 69.38\\
\hline
\\
\multicolumn{10}{l}{%
  \begin{minipage}{10.3cm}%
    \scriptsize Note: Values given are the percentage contribution.  Regions in different rows refer to the annulus receiving the flux, while columns are the annuli providing the flux.  N = north, S = south, and SE = southeast. The regions are shown in Figure \ref{fig:image}.%
  \end{minipage}%
}\\
\\
\hline\hline
\end{tabular}
\label{table:psf}
\end{center}
\end{table*}

To detect X-ray point sources unresolved by {\it Suzaku}, we use the
three {\it Chandra} pointings of the cluster, which overlap
with the northern, central, and southern {\it Suzaku} pointings. The
locations of the point sources in the field-of-view (FOV) are detected
using CIAO's  \textit{wavdetect} tool and are shown in the right panel of Figure
\ref{fig:image}. Since the point spread function (PSF) sizes of
{\it Suzaku} and {\it Chandra} are different, the extents of the point sources
reported by \textit{wavdetect} cannot be used directly to exclude
point source in the {\it Suzaku} FOV. 
We use the following procedure to determine a reliable and
conservative radius for point source exclusion. The brightest point
source within both the {\it Chandra} and
{\it Suzaku} FOV (J2000; RA: 202.603$^\circ$, DEC: $-$1.808$^\circ$)
is selected as a test case (shown with green circle in Figure \ref{fig:image} left panel). 
The source is located in a fairly faint region 
(9$^{\prime}$ away from the center of A1750 to northeast). The {\it Chandra} spectrum of the point-source
is extracted using CIAO's {\it specextract} tool and is fitted with
an absorbed power-law model with an index fixed to 1.4 (the slope 
associated with the X-ray background spectrum at 0.5$-$8 keV; e.g. \citet{hickox2006}), while the
normalization is left free. Based on the best-fit power-law index and normalization
(5.22 $\times\, 10^{-6}$ photons keV$^{-1}$ cm$^{-2}$ s$^{-1}$) obtained from the {\it Chandra} fits, a 120~ks long {\it Suzaku} XIS observation is simulated using the
\textit{xissim} tool. To assess the impact of the source flux on the measured parameters of the diffuse emission, we add the simulated source spectrum to a typical diffuse emission spectrum with 1000 net counts.  We then incrementally increase the source exclusion radius (thereby decreasing the contribution of the point source to the total emission) and examine the effect on the best-fitting parameters to the total (source plus diffuse emission) spectrum.  We find that for all exclusion radii $r > 35^{\prime\prime}$ the best-fitting parameters (kT, abundance, and normalization) are not significantly affected by the point source contribution. Since this estimate is based on the analysis of the brightest
 brightest point source in a faint region, the exclusion radius for fainter point
sources would be smaller. We note that 
since all our spectral extraction regions include at least 2000 total counts, this radius represents a conservative estimate. We therefore exclude regions with radii of 35$^{\prime\prime}$
around point sources detected by {\it Chandra} from our {\it Suzaku} analysis.

The southeast {\it Suzaku} pointing does not have an overlapping
{\it Chandra} observation. Therefore, the point sources in this region are
detected from the Suzaku data using CIAO's \textit{wavdetect} tool. The detection is
performed using {\it Suzaku}'s half-power radius of 1$^\prime$ as the
wavelet radius, as done in \citet{urban2014}. The point sources
detected with {\it Suzaku} are shown as green regions in the left panel in Figure \ref{fig:image}.

Spectra are extracted from the filtered event files in \textit{XSELECT}.
Corresponding detector redistribution function (RMF) files are constructed using the \textit{xisrmfgen} tool, while the ancillary response function (ARF) files are constructed using the \textit{xisarfgen} tool assuming a uniform surface brightness in a 20$^\prime$ radius. Cutoff-rigidity-weighted particle-induced background spectra are extracted from the NTE data for each detector using the \textit{xisnxbgen} tool. The particle induced background spectrum is subtracted from each source spectrum prior to fitting. Spectral fitting is performed in the 0.5 $-$ 7 keV energy band where the {\it Suzaku} XIS is the most sensitive. 

The cluster emission is modeled with an absorbed single temperature
thermal plasma \apec model with ATOMDB version 2.0.2
\citep{smith2001,foster2012}. \textit{XSPEC} v12.8.2 is used to
perform the spectral fits \citep{arnaud1996} with the extended
C-statistic as an estimator of the goodness of fits. We co-add front
illuminated (FI) XIS0 and XIS3 data to increase the signal-to-noise,
while the back illuminated (BI) XIS1 data are modeled simultaneously
with the front illuminated observations due to the difference in
energy responses. We adopt the solar abundance table from
\citet{andersgrevesse1989}. The galactic column density is frozen at the
Leiden/Argentine/Bonn (LAB) Galactic HI Survey value
\citep{kalberla2005} of 2.37 $\times$ 10$^{20}$ cm$^{-2}$ in our fits.

We examine the local X-ray background emission using the ROSAT All Sky
Survey ({\it RASS}) data extracted from a 1$-$2 degree annulus
surrounding the central sub-cluster's
centroid\footnote{http://heasarc.gsfc.nasa.gov/cgi-bin/Tools/xraybg/xraybg.pl}. A region 19$^\prime$ $-$
21$^\prime$ away from the central sub-cluster A1750C in the southeast pointing
is used to extract the local background (see Figure
\ref{fig:image}). The RASS spectrum is simultaneously fit with the local
background XIS FI and BI spectra using two Gaussian models for solar wind charge
exchange at 0.56 and 0.65 keV, an unabsorbed {\it apec} 
model for Local Hot Bubble (LHB) emission, and
an absorbed {\it apec} model for Galactic Halo (GH) emission
\citep{kuntz2000, bulbul2012}. The abundances of these \apec models are set to solar, while the redshifts are fixed at zero.
An absorbed power-law component
with a photon index of 1.4
is added to the model to include emission from unresolved extragalactic
sources (primarily AGN). We note that statistical uncertainties in the observed local background parameters given in this section are 1$\sigma$. The best-fit temperature of the LHB component is 0.14$_{-0.01}^{+0.03}$ keV, with a normalization of 3.22$_{-0.67}^{+3.94}\, \times 10^{-6}$ cm$^{-5}$ arcmin$^{2}$.  The best-fit temperature and normalization of the GH component is 0.69$_{-0.09}^{+0.11}$ keV and 1.79$ _{-0.43}^{+0.78}\,\times$ $\, 10^{-7}$ cm$^{-5}$ arcmin$^{2}$. The normalization for the CXB power-law component is 5.84$_{-0.63}^{+1.50}\ \times 10^{-7}$ photons keV$^{-1}$ cm$^{-2}$ s$^{-1}$ arcmin$^{-2}$ at 1 keV, corresponding to a CXB flux of (1.15 $\pm\ 0.30)\ \times\ 10^{-11}$ ergs s$^{-1}$ cm$^{-2}$ deg$^{-2}$. The flux (6.22 $\pm$ 0.16)~$\times\ 10^{-12}$ ergs s$^{-1}$ cm$^{-2}$ deg$^{-2}$ in the 0.5$-$2 keV band is in agreement with the value (7.7$\pm$0.4~$\times\ 10^{-12}$ ergs s$^{-1}$ cm$^{-2}$ deg$^{-2}$) reported by \citep{bautz2009}.

\section{Systematic Errors on X-ray Observables}
\label{sec:syst}

In studies of low surface brightness emission, it is crucial to estimate the contribution of various systematic
uncertainties, particularly those related to background modeling, to the
total error budget. We consider the following potential sources of systematic
error in our analysis; i) uncertainties due to stray light contamination and the large size of the PSF of
{\it Suzaku}'s mirrors; ii) uncertainties due to intrinsic spatial
variations in the local 
soft background; iii) systematics associated with the NXB; vi) uncertainties due to the intrinsic spatial
variation of unresolved point sources.

\subsection{Scattered Light due to Large PSF}
\label{sec:bkgsyst}

Due to {\it Suzaku}'s relatively large PSF, some X-ray photons that
originate from one particular region on the sky may be detected
elsewhere on the detector. 
The PSF spreading in each direction is calculated by
generating simulated event files using the ray-tracing simulator
\textit{xissim} \citep{serlemitsos2007}. {\it Chandra} X-ray images of each
annular sector (shown in Figure \ref{fig:image} left panel) and the best-fit
spectral models obtained from the {\it Suzaku} observations are used to
simulate event files with 1$\times\ 10^{6}$ photons. 
The fraction of photons that are spread into the surrounding annuli is calculated for each XIS detector and annulus sector. 
Relative contributions are weighted by the effective area at 1.5 keV of
each detector to calculate the overall percentage contribution (given in Table \ref{table:psf}). We find that the majority of photons
originating in an annulus on the sky are detected in the same annulus (except region N1)
on the detector. Up to 15\%
of the photons may be detected in surrounding annuli. However, the
percentage fraction of photons that scatter into the outermost annuli at $R_{200}$\footnote{The overdensity radius $R_{200}$ is defined as the radius within which the average matter density of the cluster is 
200 times the critical density of the Universe at the cluster redshift.}  to the north
and southeast from the bright cores is small ($<$1\%). 
These results are consistent with the photon fractions reported in \citet{bautz2009} and \citet{walker2012a}.

Considering the shallow temperature distribution of A1750 measured by {\it Chandra} observations, the PSF is
expected to have a minimal effect on the measurements of temperature in
the outermost regions. To estimate the effect of PSF spreading on our
temperature and normalization measurements, we jointly fit
the spectra of sectors with \apec models, with the normalizations
scaled according to the fractions listed in Table \ref{table:psf}. In
all cases,  the change in best-fit parameter values due to scattered flux from other annuli is significantly less than the statistical errors on the measured observables (see Table \ref{table:1tfits}).

\begin{table}
\def\arraystretch{1.7}
\begin{center}
\caption{Systematic Soft, Cosmic, and Particle X-ray Background Uncertainties }
\begin{tabular}{lcccc}
\hline\hline
Pointing 		&Region 1 	&Region 2  	& Region 3 	& Region 4\\
	 	\hline
North		& 1	 		& 4  			& 8 			& 19	\\
Southeast 	& 0.7			& 3.4	 		& 25			& -	\\
South		& 4			& 20			& - 			& -  \\
\hline
\\
\multicolumn{5}{l}{%
  \begin{minipage}{7.3cm}%
    \scriptsize Note: Values are the percentile systematic uncertainties on temperature for each region due to soft Galactic, cosmic X-ray, and the particle background.%
  \end{minipage}%
}\\
\\
\hline\hline
\end{tabular}
\label{table:bkgfluc}
\end{center}
\end{table}
%

\subsection{Systematics Related to Soft, Cosmic, and Particle X-ray Background} 
\label{sec:systcxb}

To model the soft X-ray foreground and cosmic X-ray background, we
jointly fit \rosat RASS data with local XIS background spectra, as
described in detail in Section \ref{sec:suzanalysis}. We find that the
local X-ray background is consistent with the RASS data. However,
spatial variations in the background level can introduce additional
systematic uncertainties on X-ray observables. To estimate the effect
of these uncertainties, we perform 10,000 Monte Carlo realizations of the background
model. The model parameters are allowed to vary
simultaneously within their 1$\sigma$ uncertainty ranges obtained from
the joint RASS-local background fit. A variation of up to $\sim$3.6\% of the NXB
level is also taken into account \citep{tawa2008}. The percent
systematic uncertainty contributions due to the variance in cosmic,
local, and particle background on the temperature estimates are given
in Table \ref{table:bkgfluc}. We find that the effect on temperature
and normalization is negligible ($\sim$ 1\%) and smaller than the
statistical uncertainties in the inner regions (shown in Table
\ref{table:1tfits}), while it can be as large as 25\% in the outskirts
near $R_{200}$. These uncertainties are
included in the total error budget in our analysis by adding them in
quadrature.

\begin{table}
\def\arraystretch{1.7}
\begin{center}
\caption{Estimated 1$\sigma$ Fluctuations in the CXB level due to unresolved point sources in the {\it Suzaku} FOV in units of $10^{-12}$ ergs cm$^{-2}$ s$^{-1}$ deg$^{-2}$.}
\begin{tabular}{lcccc}
\hline\hline
Pointing 	 	& Region 1 	& Region 2  	& Region 3 	& Region 4\\
		 \hline
North		& 17.6 		& 10.2 		& 7.9 		& 4.4	 \\
Southeast 	& 11.1 		& 6.8			& 4.3 		& -	\\
South		& 9.9 		& 4.7 		& - 			& -  \\
\hline\hline
\end{tabular}
\label{table:cxbfluc}
\end{center}
\end{table}
%


\subsection{Systematics Related to Cosmic X-ray Background} 
\label{subsec:systbkg}

The intrinsic variations in the unresolved CXB component can be an important source of uncertainty
in the analyses of cluster outskirts with {\it Suzaku}. To estimate the
magnitude of this component, we follow a similar approach to that
described in \citet{walker2012}. The {\it Suzaku} data alone allow us to detect point sources to a limiting flux of 1.3 $\times\ 10^{-14}$ ergs cm$^{-2}$ s$^{-1}$ deg$^{-2}$ in our observations. 

The contribution of unresolved point sources to the total flux in ergs cm$^{-2}$ s$^{-1}$ deg$^{-2}$ can be estimated as \citep{moretti2003}:
\begin{equation}
F_{CXB} = (2.18\pm0.13)\times 10^{-11}  -\int_{S_{excl}}^{S_{max}}\frac{dN}{dS}\times S\ dS.
\label{eqn:cxbflux}
\end{equation}
The source flux distribution in the 2$-$10 keV band  is described by the analytical function
\begin{equation}
N(>S) = N_{0} \left[\frac{(2\times10^{-15})^{\alpha}}{S^{\alpha}+S_{0}^{\alpha-\beta}}\right] \rm{erg\ cm^{-2}\ s^{-1}},
\label{eqn:psflux}
\end{equation}

\noindent where $N_{0}=5300^{+2850}_{-1400}$, $S_{0}=(4.5^{+3.7}_{-1.7})\times 10^{-15}$, $\alpha=1.57^{+0.10}_{-0.18}$, and $\beta=0.44^{+0.12}_{-0.13}$. We then integrate Equation \ref{eqn:cxbflux} from a lower limit of $S_{excl} = 1.3 \times\, 10^{-14}$ ergs cm$^{-2}$ s$^{-1}$ (the flux of the faintest source in our FOV) up to the upper limit of $S_{max}=8.0\times10^{-12}$ ergs cm$^{-2}$ s$^{-1}$ \citep{moretti2003}. The integration gives an unresolved 2 $-$ 10 keV flux of (1.20 $\pm\ 0.43)\, \times \ 10^{-11}$ ergs cm$^{-2}$ s$^{-1}$ deg$^{-2}$.

The expected deviation in the CXB level due to unresolved point sources is 
\begin{equation}
\sigma_{B}^{2}= \frac{1}{\Omega}\int_{0}^{S_{excl}} \frac{dN}{dS}\times S^{2}\ dS,
\label{eqn:fluc}
\end{equation}

\noindent where $\Omega$ is the solid angle \citep{bautz2009}. Using the power-law relation (given in Equation \ref{eqn:psflux}) in Equation \ref{eqn:fluc}, we calculate the 1$\sigma $~RMS fluctuations in the CXB (given in Table \ref{table:cxbfluc}). We find that the variation is 4.3 $\times\ 10^{-12}$ erg cm$^{-2}$ s$^{-1}$ deg$^{-2}$ in the faintest outermost SE3 region, which extends to $R_{200}$. These estimates are consistent with the values reported by \citet{bautz2009} and \citet{hoshino2010}. The 1$\sigma$ uncertainty on the measured CXB (from joint RASS and local background fits) is comparable to the expectation value of fluctuations on the CXB brightness calculated here. We include this variation in our Monte-Carlo Markov realizations of the X-ray background to account for the CXB variation (as described in detail in Section \ref{sec:systcxb}). The final systematic errors on the observed quantities were added in quadrature.

\begin{figure}
\centering
\hspace{-4mm}\includegraphics[width=8cm, angle=0]{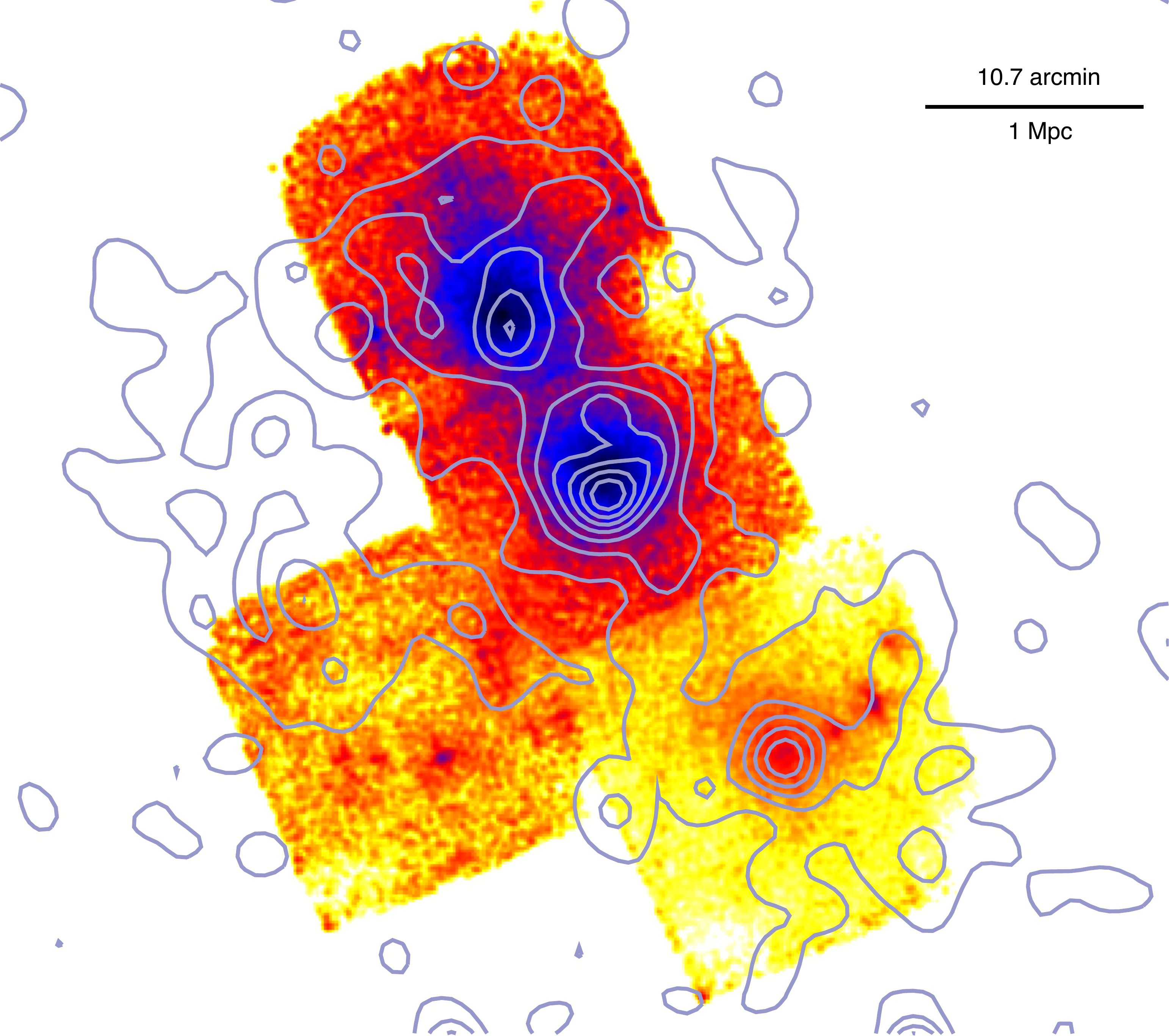}
\vspace{3mm}
\caption{ {\it Suzaku} X-ray image of the A1750 merger system with contours
  from the smoothed optical red light distribution overplotted in blue.}
\vspace{3mm}
\label{fig:rsmap}
\end{figure}
%


\section{Optical Properties and Merger Dynamics}
\label{sec:resultsopt}

In this section, we describe our search for substructure in the optical data and
use them to further constrain the dynamical state and merger history
of A1750 and determine whether the subclusters are bound to each other.

\subsection{The Red Sequence Galaxy Population}

In Figure \ref{fig:rsmap} we overplot the spatial density of red sequence (RS) galaxies on the sky on the {\it Suzaku} X-ray image. This map is created from a cloud-in-cell interpolation of the spatial distribution of red 
sequence selected galaxies on the sky, where galaxies are weighted by their r-band magnitudes 
(brighter galaxies weighted more heavily). The resulting map of the surface density of red sequence 
light traces the collisionless galaxy component of the system. We then
applied a broad gaussian smoothing kernel of 54$^{\prime\prime}$ to
generate the contours shown in the image.

The peaks observed in the red light distribution roughly align with the peaks in
the X-ray emission. We note that we do not find strong evidence for extended filaments along the axis of the aligned clumps in RS light, thus no large-size groups are detected along the filament direction to the North. The lack of evidence indicates that the filaments do not contain significant 
large group-like structures with a detectable red sequence population.

\subsection{Spectroscopic Properties of Cluster Member Galaxies}
\label{sec:optspec}

The large sample of spectroscopic redshifts available in the A1750 field provides 
an opportunity to investigate the dynamical state of cluster member galaxies. We first 
characterize the cluster member dynamics for the entire 
system by making an initial selection of cluster members that are within a projected physical 
radius of 1 Mpc of the centroid of the X-ray emission of each subcluster (this is approximately equal to the 
region covered by our Hectospec observations), and which have redshifts in the 
interval 0.07 $< z <$ 0.1. We then use the bi-weight location and scale estimators 
\citep{beers1990} as the starting guess for the median and dispersion of cluster member 
velocities. We then iterate this process, rejecting galaxies with redshifts more than 
3-$\sigma$ away from the median until the redshift sample converges. This results in 
an estimate of the velocity dispersion for the entire system of $\sigma_{v} = 780 \pm 30$ 
km s$^{-1}$, and a median redshift of $\bar{z} = 0.0861 \pm 0.0002$ (a recession velocity, 
$\bar{v} = 23780 \pm 50$ km s$^{-1}$) based on 243 cluster member redshifts.

\begin{figure}
\centering
\hspace{-4mm}\includegraphics[width=8cm, angle=0]{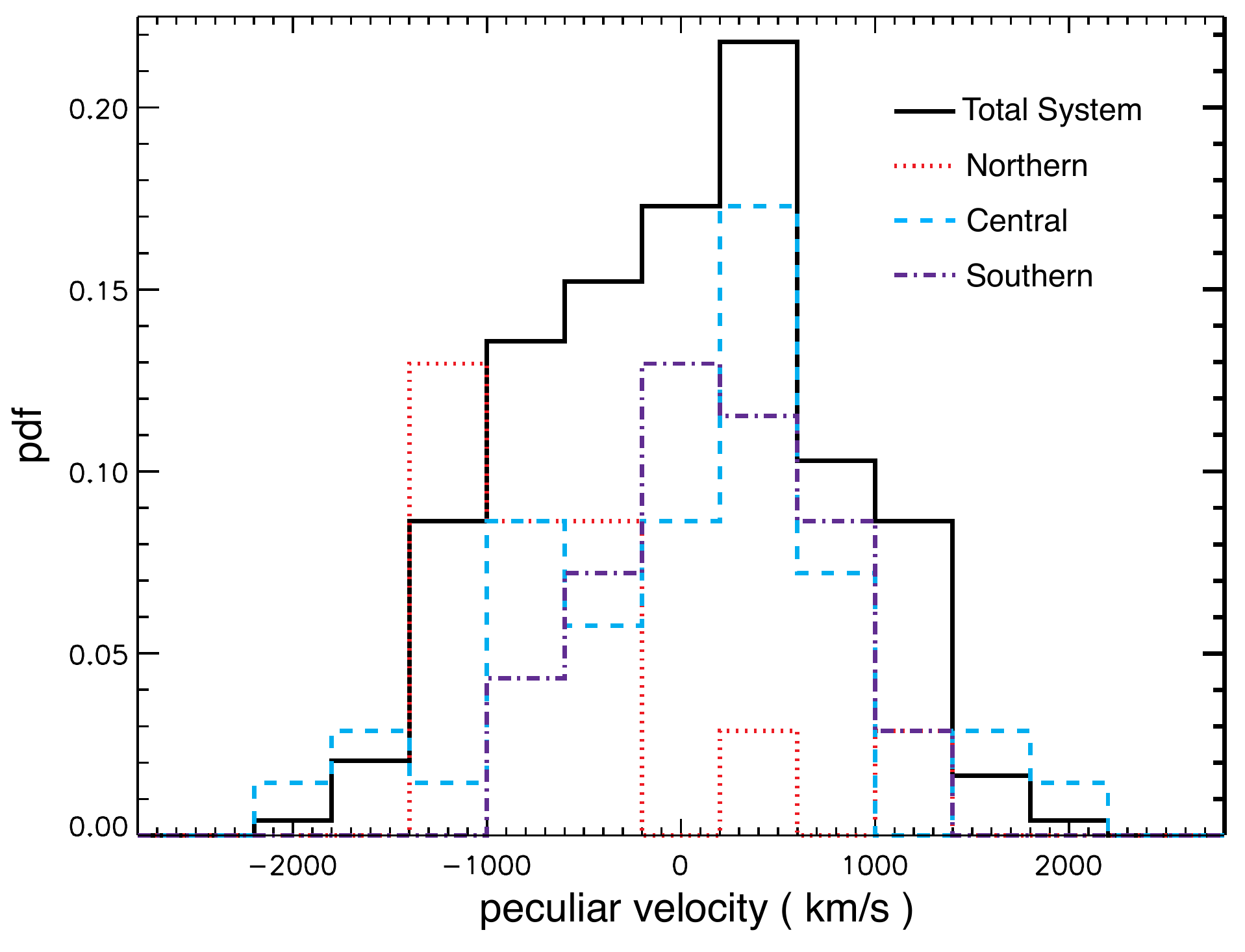}
\vspace{3mm}
\caption{Histograms of the velocity distribution of galaxies that sample the sub-clusters and the merger superstructure. Values of the biweight median and variance are listed in Table \ref{table:memberdynamics}. }
\vspace{3mm}
\label{fig:vdist}
\end{figure}
\begin{table}
\def\arraystretch{1.5}
\begin{center}
\caption{Cluster Member Dynamics\label{tab:member dynamics}}
\begin{tabular}{lcccc}
\hline\hline\\
Region	& $N_{mem}$ 	& $z_{med}$	& $\sigma_{1D}$	& $\Delta v^{a}$\\
		&				&			&  (km s$^{-1}$) 	&  (km s$^{-1}$)  \\
	 \hline
Total System   & 243 & 0.0861$\pm$0.0028 & 780$\pm$30  & 0  \\
A1750N   &  25 & 0.0832$\pm$0.0006 & 750$\pm$160 & -810  \\
A1750C  &  40 & 0.0864$\pm$0.0004 & 835$\pm$120 & 90  \\
A1750S   &  33 & 0.0868$\pm$0.0019 & 532$\pm$60  & 200   \\	
\hline
\\
\multicolumn{5}{l}{%
  \begin{minipage}{7.3cm}%
    \scriptsize $^{a}$ ~Peculiar velocity relative to the recession velocity of the entire merging system.%
  \end{minipage}%
}\\
\\
\hline\hline
\end{tabular}
\label{table:memberdynamics}
\end{center}
\end{table}

Given that we have hundreds of spectroscopic cluster members, we can also test for  
line-of-sight velocity differences between the three individual X-ray sub-clusters. We define 
subsets of spectra that originate from galaxies located in three non-overlapping 3$^{\prime}$
radius circular regions on the sky that are centered on the X-ray
peaks of each of the three distinct sub-clusters. For each of these regions, we use all of the cluster member galaxies 
that satisfy the $\pm$3-$\sigma$ velocity range for the total cluster system from above, and compute 
the bi-weight location and scale estimates of the median and dispersion in the galaxy velocities. 
These regions extend out radially $\sim$300 kpc from each X-ray centroid, and therefore only 
include a relatively small fraction of the full sample of 243 cluster member spectra (between 
25$-$40 cluster members per subregion). 

The resulting kinematics estimates are given in Table~\ref{table:memberdynamics}; the central and 
southern X-ray clumps have redshifts that are similar to the median for the 
total system, but the northern clump is blue-shifted, with a peculiar velocity of $-810$ km 
s$^{-1}$ (see Figure \ref{fig:vdist}). The observed peculiar velocities imply that any relative motion between the central and 
southern clumps is in the plane of the sky, while the northern clump is moving at least 
partly along a vector that is normal to the sky.

The velocity dispersion of galaxies within the total structure is not larger than the velocity dispersion of the individual clumps, indicating that the system is unrelaxed. The individual subclusters haven't begun to virialize into the final larger cluster, i.e. the total mass of all three clumps is $\sim\ 10^{15}\ M_{\odot}$, and 840 km s$^{-1}$ is well below the velocity dispersion of a virialized structure of that mass. The shape of the velocity distributions within the different subcluster regions (plotted in Figure \ref{fig:vdist}), suggest that northern and central subclusters are less well-structured (with asymmetric velocity dispersion profiles) than that of the southern subcluster. This could be due to some degree of interaction between the central and northern subclusters, while the southern subcluster may still be infalling (i.e., has not started tidally interacting with the other systems).

We further calculated the implied virial masses of individual
subclusters based on the velocity dispersions. Using the \citet{evrard2008} scaling relations, the virial masses of A1750N, A1750C, and A1750S are 4.6$_{-2.4}^{+3.6} \times10^{14}$  M$_\odot$, 6.4$_{-2.4}^{+3.1}\times10^{14}$  M$_\odot$, and 1.7$_{-0.5}^{+0.6}\times 10^{14}$  M$_\odot$, respectively. These masses are consistent with total masses of each subcluster obtained from X-ray observations (see Section \ref{sec:mass} for detailed calculations).

\begin{table}
\def\arraystretch{1.5}
\begin{center}
\caption{Best-Fit Parameters for the Bound Incoming Solutions of the A1750N-A1750C System \label{tab:bound}}
\begin{tabular}{cccccc}
\hline\hline\\
$\chi$&$\alpha$ &$R$ &$R_{\rm m}$ &$V$ & $P$ \\
	(rad) & (degrees) & (kpc) & (kpc) & ($\rm km~s^{-1}$) &($\%$)\\
\hline
4.542 & 71.73 & 2966.7 & 5072.2 & 931.2  & 22 \\ 
5.319 & 23.86 & 1016.9 & 4727.4 & 2186.2 & 78 \\  
\hline
\\
\multicolumn{6}{l}{%
  \begin{minipage}{7.3cm}%
    \scriptsize Note: ~Columns list best-fit values for $\chi$ and $\alpha$
for the bound solutions of the dynamical model, and the corresponding values for
$R$, $R_{\rm m}$, $V$, and the probability of each solution.%
  \end{minipage}%
}\\
\\
\hline\hline
\end{tabular}
\end{center}
\end{table}
%


\subsection{A dynamical model for the A1750N-A1750C System}
\label{sec:dynamics}

We apply a dynamical model introduced by
\citet{1982Beers} and \citet{as2015} to evaluate the dynamical state 
of the subclusters A1750N and A1750C. This model allows us to estimate 
the most likely angle between the merger axis and the plane of the
sky.

The equations of motion take two different forms, depending on whether
the subclusters are gravitationally bound or not.
For the case where they are gravitationally bound, 
we parameterize the equations of motion in the following form:
\begin{eqnarray}
&&R=\frac{R_{\rm m}}{2}(1 - \cos \chi),\label{eq:r_bound} \\
&&t=\left(\frac{R_{\rm m}^3}{8GM}\right)^{1/2}(\chi - \sin\chi), \\
&&V=\left(\frac{2GM}{R_{\rm m}}\right)^{1/2}\frac{\sin\chi}{(1 -
  \cos\chi)},
\end{eqnarray}
where $R_{\rm m}$ is the subclusters' separation at the moment of maximum
expansion, $M$ is the system's total mass, and $\chi$ is the variable
used to parametrize Friedmann's equation, also know as development
angle. For the 
case of not gravitationally bound subclusters, 
the equations are parametrized as:
\begin{eqnarray}
&&R=\frac{GM}{V_\infty^2}(\cosh\chi - 1), \\
&&t=\frac{GM}{V_\infty^3}(\sinh\chi - \chi), \\
&&V=V_\infty\frac{\sinh\chi}{(\cosh\chi - 1)},
\end{eqnarray}
where $V_\infty$ is the velocity of expansion at the asymptotic limit.
$V_r$, the radial velocity difference, and $R_p$, the projected distance, 
are related to the parameters of the equations by
\begin{eqnarray}
V_{\rm r} = V~\sin\alpha, ~ R_{\rm p} = R~\cos\alpha,\label{eq:v_r_alpha}
\end{eqnarray}

\noindent  where $\alpha$ is the projection angle of the system with
respect to the plane of the sky.

The virial mass of this subclusters is $M = (7.2 \pm 1.0) \times 10^{14}
~M_\odot$ (sum of the masses of both subclusters within $R_{200}$ --- uncertainties are quoted here as 68\% confidence intervals; see Section \ref{sec:mass} for detailed calculations)
derived from \citet{bulbul2010} ICM models .  
We assume that the subclusters' velocities are the median velocities of
their galaxies. The projected distance on the plane of
the sky between the X-ray center of each subcluster is $R_{\rm p}$ =
0.93 Mpc. The difference of the median redshifts
of these subclusters yields a radial velocity difference of 
$V_r = 884 \pm 199 {\rm~km~s^{-1}}$. 
By setting $t$ = 12.4 Gyr, the age of the
Universe at the mean redshift of these subclusters ($z=0.0848$), we close the system of equation. 
The parametric equations are then solved via an iterative procedure, 
which computes the radial velocity difference $V_{\rm r}$ 
for each projection angle $\alpha$.

Using simple energy considerations, we determine the limits of the bound
solutions:
\begin{eqnarray}
V_{\rm r}^2  R_{\rm p} \leq 2GM~{\sin^2\alpha \cos\alpha}.
\label{eq:limit_bound_solutions}
\end{eqnarray}

\begin{figure}[h!]
\centerline{\includegraphics[width=.5\textwidth]{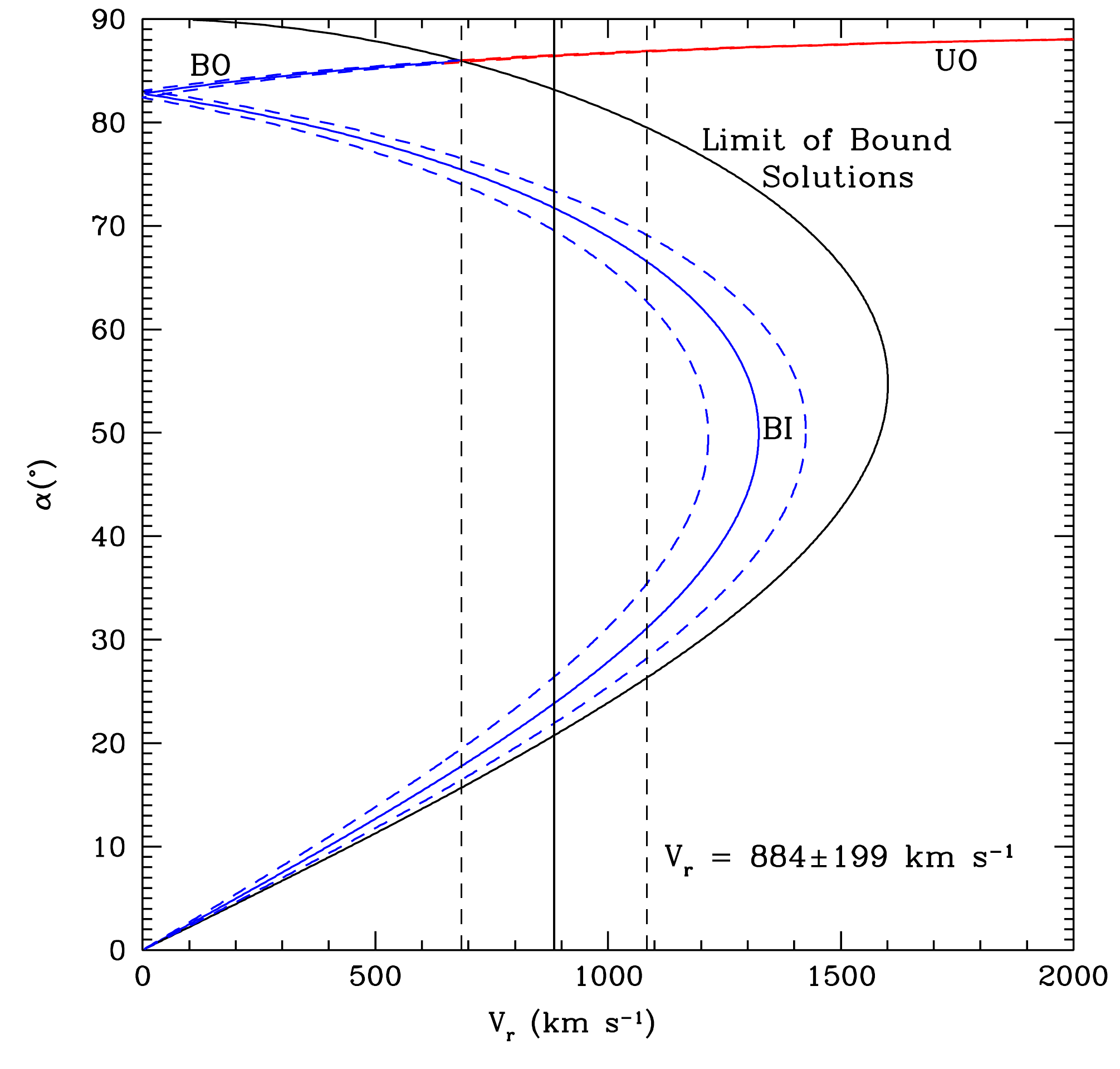}}
\caption{Projection angle ($\alpha$) as a function of the radial
velocity difference ($V_{\rm r}$) between the subclusters. UO, BI, and BO stand for Unbound Outgoing, Bound
Incoming, and Bound Outgoing solutions.
Solid red and  blue lines correspond to unbound and bound solutions,
respectively. The vertical solid line corresponds to the radial velocity
difference between the median velocities of the galaxies in each subcluster. Dashed lines correspond to
68$\%$ confidence ranges.
\label{fig:alpha_vr}}
\end{figure}

Figure \ref{fig:alpha_vr} presents the projection angle ($\alpha$) as a function of the radial
velocity difference ($V_{\rm r}$) between the subclusters. 
The uncertainties in the measured radial velocity and 
mass of the subclusters lead to a range in the solutions for the
projection angles ($\alpha_{\rm inf}$ and $\alpha_{\rm sup}$). 
We compute the relative probabilities of these solutions by:
\begin{eqnarray}
p_i = \int_{\alpha_{{\rm inf},i}}^{\alpha_{{\rm sup},i}} \cos\alpha ~d\alpha,
\end{eqnarray}
where each solution is represented by the index $i$. We then normalize
the probabilities by $P_{i} = p_{i}/(\sum_i p_i)$.

\begin{table}
\def\arraystretch{1.5}
\begin{center}
\caption{Best-Fit Parameters for the Unbound Outgoing Solution of the Dynamical Model of the A1750N-A1750C System.\label{tab:unbound}}
\begin{tabular}{cccccc}
\hline\hline\\
$\chi$&$\alpha$ &$R$ &$R_{\rm m}$ &$V$ & $P$ \\
	(rad) &(degrees) &(kpc) &(kpc) &($\rm km~s^{-1}$) &($\%$)\\
\hline
1.692 & 86.45 & 15025.5 & 886.0 & 610.4 & 0.02\\
\hline
\\
\multicolumn{6}{l}{%
  \begin{minipage}{7.3cm}%
    \scriptsize Note: ~Columns present best fit for the $\chi$ and $\alpha$
for the unbound solution, and the corresponding values for
$R$, $V$, $V_\infty$, and probability of this solution.%
  \end{minipage}%
}\\
\\
\hline\hline
\end{tabular}
\end{center}
\end{table}
\begin{table}
\def\arraystretch{1.5}
\begin{center}
\caption{Best Fit Parameters for the Bound Solutions of the A1750C-A1750S System\label{tab:bound2}}
\begin{tabular}{ccccccc}
\hline\hline\\
$\chi$&$\alpha$ &$R$ &$R_{\rm m}$ &$V$ & $P$ & Relative\\
	(rad) & (degrees) & (kpc) & (kpc) & ($\rm km~s^{-1}$) &($\%$) &  Motion\\
\hline
2.871 & 77.68 & 7871.0 & 8017.1 & 113.0  & 10 & Outgoing \\ 
3.385 & 74.76 & 6390.7 & 6486.5 & 114.4  & 15 & Incoming \\ 
4.987 &  4.39 & 1684.9 & 4620.9 & 1442.4 & 75 & Incoming \\ 
\hline
\\
\multicolumn{7}{l}{%
  \begin{minipage}{7.3cm}%
    \scriptsize Note:  ~Columns list best-fit values for $\chi$ and $\alpha$
for the bound solutions of the dynamical model, and the corresponding values for
$R$, $R_{\rm m}$, $V$, and the probability of each solution.%
  \end{minipage}%
}\\
\\
\hline\hline
\end{tabular}
\end{center}
\end{table}

Solving the parametric equations we obtain two bound solutions and one unbound solution. 
For the case of the bound solutions, the subclusters are either approaching each other at
931 km s$^{-1}$ (22$\%$ probability) or at 2186 km s$^{-1}$
(78$\%$ probability). The former solution corresponds to a collision in less than 3.1 Gyr, given their 
separation of $\sim$ 2.97 Mpc. The latter corresponds to a collision in less than 460 Myr, given their separation 
of $\sim$  1017 kpc. The unbound solution (0.02$\%$ probability) corresponds to a separation of $\sim$ 15 Mpc.
The parameters of these solutions are presented in Tables \ref{tab:bound} and
\ref{tab:unbound}. Given its very low probability, the unbound solution can
be neglected, while the bound solution in which the separation
between the clusters is $\sim$ 1017 kpc is highly favored ($78\%$
probability).

As mentioned in \citet{as2015}, 
the method to determine the dynamical state of a system
of clusters from \citet{1982Beers} assumes a purely radial infall. 
Also, the way the probabilities are computed, by integrating over the
angles determined by the uncertainties on the mass of the system,  
favors small angle solutions. 
Therefore, the probabilities for the solutions should be
treated with caution, as we have no information about the angular momentum
of this subclusters.

\subsection{A dynamical model for the A1750C-A1750S System}

Now, we apply the same procedure to determine the dynamical state of
the pair A1750C-A1750S. Using the virial mass estimated from 
the velocity dispersion of the galaxies in the southern subcluster, 
the total mass of this system is $M = (6.5 \pm 1.7) \times 10^{14}
~M_\odot$ (uncertainties are quoted here at the 68\% confidence
level). 
The difference between the median redshifts of 
these subclusters yields a radial velocity difference of 
$V_r = 110 \pm 123 {\rm~km~s^{-1}}$. Solving the system of parametric
equations (Equations (\ref{eq:r_bound}) -- (\ref{eq:v_r_alpha})) 
yields the results presented in Table \ref{tab:bound2}, with
A1750C-A1750S being bound in all solutions. The most likely solution
(75\% probability) indicates that the merger is happening very close to the plane 
of the sky ($\alpha = 4$ degrees), also supporting the scenario in
which all three subclusters are merging along a cosmic filament.

\section{The Observed ICM Properties}
\label{sec:resultsxray}

We extract spectra in concentric annular sectors along the north
(filament), south (filament), and southeast (off-filament) directions
from the regions shown in Figure \ref{fig:image}. Each spectral extraction region is
selected to include at least 2000 net source counts. The LHB+CXB+GH 
components are fixed to the values determined from fits to local
background and RASS data as described in Section
\ref{sec:suzanalysis}. We stress that the systematic errors are included as explained in Section 
\ref{sec:syst}. The {\it Suzaku} spectra are fitted using an absorbed single
temperature (1T) \apec model with free temperature, abundance, and normalization. 


\subsection{Filament Direction}
\label{sec:fila}

We first examine the {\it Suzaku} spectra extracted along the north direction
starting from the center of A1750N. The spectra are extracted from
four consecutive annular sectors;
0$^\prime$$-$2.5$^\prime$, 2.5$^\prime$$-$5$^\prime$,
5.0$^\prime$$-$7.5$^\prime$, and 7.5$^\prime$$-$12.5$^\prime$. The total source counts in the
co-added FI observations in regions N1, N2, N3, and N4 are 3300, 4400, 2700,
and 3300, respectively. 
The BI spectra in the same regions have total source counts of
2400, 3300, 2100, and 2600. Both FI and BI spectra of the outermost 7.5$^\prime$
$-$ 12.5$^\prime$ region are dominated by the NXB background at $>$6 keV,
thus this band is excluded from further analysis.

\begin{table}
\def\arraystretch{1.4}
\begin{center}
\caption{Best-fit parameters of the 1T \apec model }
\begin{tabular}{lcccc}
\hline\hline\\
	 
Region	& $kT$ 		& Abund	 		&$ \mathcal{N}$		& C-Stat\\
	 	& (keV) 	& (A$_{\odot}$)	& ($10^{-6}$ cm$^{-5}$)	& (dof)\\
	 \\
	 \hline
\\
N$_{1}$	& 3.33 $^{+0.17}_{-0.14}$ & 0.28 $\pm$ 0.5 	& 110.18 $\pm$ 2.95	& 178.10  (177) \\
N$_{2}$	& 2.80 $^{+0.16}_{-0.20}$	& 0.15 $\pm$ 0.4	& 48.87 $\pm$ 1.70	& 183.31 (244)	 \\
N$_{3}$	& 1.98 $\pm$ 0.18		& 0.2$^{*}$		& 17.55 $\pm$ 0.77	& 272.86 (165)	\\
N$_{4}$	& 1.61 $\pm$ 0.30		& 0.2$^{*}$		& 6.08 $\pm$ 1.09	& 322.35 (241)	\\
\\
S$_{1}$	& 2.61 $\pm$ 0.21		& 0.19 $\pm$ 0.08	& 22.28 $\pm$ 1.79	& 148.68 (144)	\\
S$_{2}$	& 2.04 $^{+0.30}_{-0.41}$ & 0.20 $_{-0.14}^{+0.24}$ & 2.76 $\pm$ 1.15 & 201.25 (187)\\
\\
SE$_{1}$	& 4.72  $^{+0.17}_{-0.13}$& 0.31 $\pm$ 0.01	& 154.79 $\pm$ 5.33	& 712.29 (746) \\
SE$_{2}$	& 4.83 $\pm$ 0.40		& 0.2$^{*}$		 & 31.93 $\pm$ 1.44 	& 776.43 (747)\\
SE$_{3}$	& 2.47 $_{-0.68}^{+0.75}$	& 0.2$^{*}$		& 4.26 $\pm$ 1.10	& 495.63 (473)\\
\\
\hline
\\
\multicolumn{5}{l}{%
  \begin{minipage}{7.3cm}%
    \scriptsize $^{*}$ indicates the fixed parameters%
  \end{minipage}%
}\\
\\
\hline\hline
\end{tabular}
\label{table:1tfits}
\end{center}
\end{table}

To investigate the nature of the gas along the filament, we
first fit the FI and BI spectra simultaneously with a 1T \apec model. 
The parameters of the FI and BI spectral models are tied to
each other. The abundances are only constrained by the observations in regions N1
and N2. The
best-fit temperatures are 3.33$^{+0.17}_{-0.14}$ keV and
2.80$^{+0.16}_{-0.20}$ keV, respectively. A 1T \apec model produces an
acceptable fit to the spectra of the innermost two regions. Adding an additional \apec model does not
significantly improve the fits for these regions. The model parameters are given in Table
\ref{table:1tfits}.

\begin{table*}
\def\arraystretch{1.7}
\begin{center}
\caption{The best-fit parameters of the 2T \apec model in regions N3 and N4 in the 0.5-7 keV energy band}
\begin{tabular}{lccccc}
\hline\hline\\
Region	& $kT_{1}$	& Abund	&$ \mathcal{N}_{1}$		& $kT_{2}$ 		&$ \mathcal{N}_{2}$	\\
	 	& (keV) 		& A$_{\odot}$	&($10^{-6}$ cm$^{-5}$)	& (keV) 	& ($10^{-6}$ cm$^{-5}$) \\
	 \\
	 \hline
	 \\
N$_{3}$	& 3.24$_{-0.55}^{+1.40}$		& 0.1$^{\star}$ & 10.32$\pm$2.05	& 1.01$^{+0.13}_{-0.07}$ 	& 9.19$_{-1.82}^{+2.42}$ 	\\
N$_{3}$	& 2.93$_{-0.40}^{+0.57}$		& 0.2$^{\star}$ & 11.76$\pm$1.66	& 0.99$\pm$0.07 	& 4.87$_{-1.04}^{+1.56}$ 	\\
N$_{3}$	& 2.93$_{-0.37}^{+0.46}$		& 0.3$^{\star}$ & 11.75$\pm$1.53	& 0.95$\pm$0.08 	& 3.29$_{-0.68}^{+0.92}$  	\\

N$_{4}$	& 1.95$_{-0.39}^{+0.62}$ 		& 0.1$^{\star}$ & 4.81$_{-1.72}^{+1.09}$	& 0.79$^{+0.19}_{-0.10}$ & 2.98$\pm0.71$ \\
N$_{4}$	& 2.12$_{-0.37}^{+0.50}$ 		& 0.2$^{\star}$ & 4.53$_{-1.01}^{+1.05}$	& 0.81$\pm$0.12 & 1.74$\pm0.36$ \\
N$_{4}$	& 2.29$_{-0.40}^{+0.50}$ 		& 0.3$^{\star}$ & 4.21$\pm$ 1.33		& 0.80$^{+0.12}_{-0.08}$  & 1.28$\pm$ 0.25 \\

\hline
\multicolumn{6}{l}{%
  \begin{minipage}{7.3cm}%
    \scriptsize $^{\star}$ Values held constant.%
  \end{minipage}%
}\\
\hline\hline
\end{tabular}
\label{table:2tfits}
\end{center}
\end{table*}
\begin{figure}
\centering
\vspace{3mm}
\hspace{-4mm}\includegraphics[width=8.6cm, angle=0]{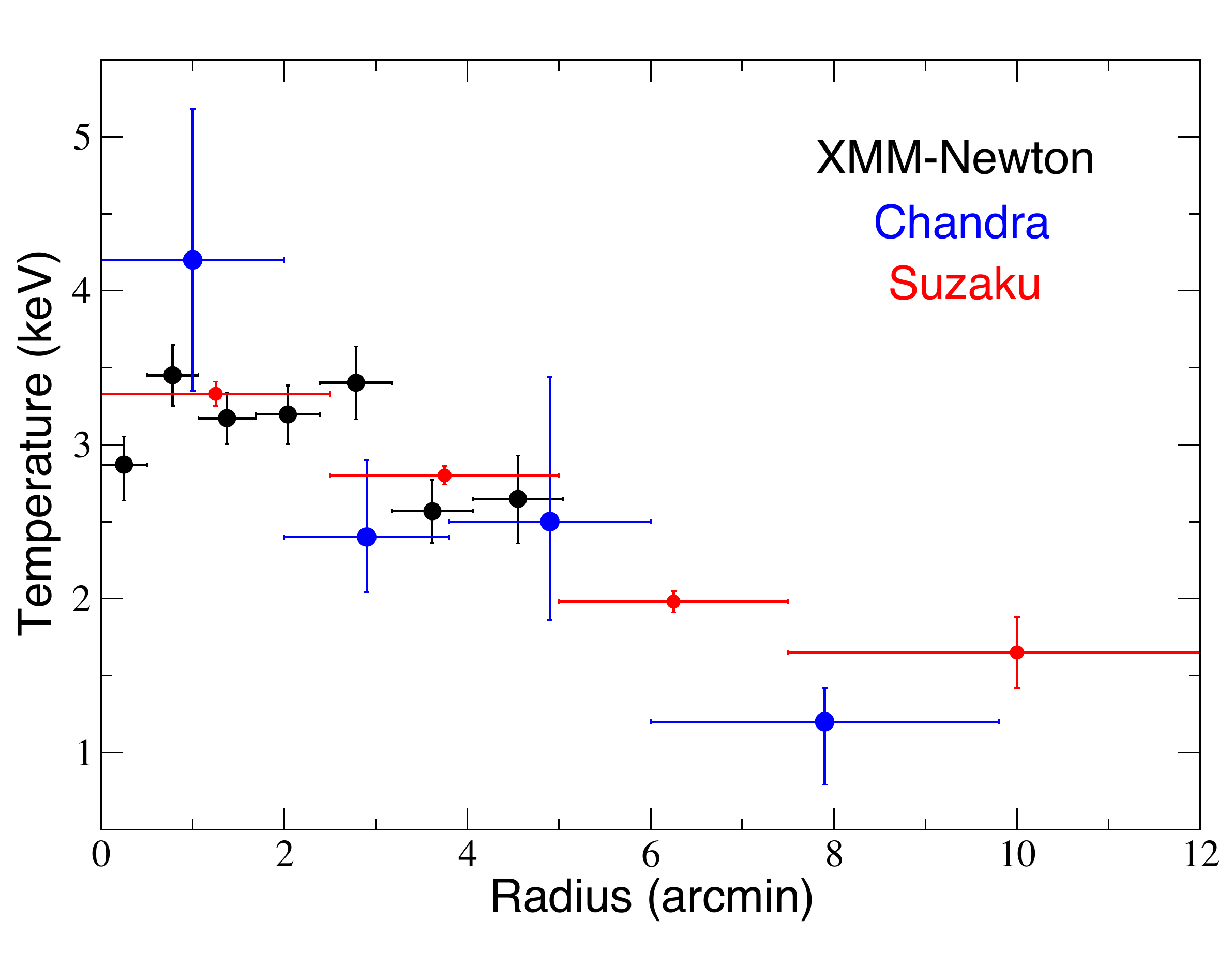}
\caption{A comparison of projected temperatures of A1750 to the north direction obtained from {\it Suzaku} (in red), {\it Chandra} (in blue), and {\it XMM-Newton} observations (in black; \citet{belsole2004}). The 1$\sigma$ error bars of {\it XMM-Newton} and {\it Suzaku} temperatures include systematic and statistical uncertainties. Temperatures reported by three satellites are in a good agreement. We are able to extend the gas temperature measurements out to 0.9 $R_{200}$ of A1750N (R$_{200}\, \sim\, 14^\prime$) and $R_{200}$ of A1750C (R$_{200}\, \sim\, 16^\prime$) clusters.}
\label{fig:projtprof}
\end{figure}

A 1T \apec model produces 
 best-fit temperatures of 1.98$\pm$0.18 keV and
 1.61$\pm$0.30 keV in regions N3 and N4, respectively. 
Abundances are not constrained; we therefore assume an abundance of
0.2A$_{\odot}$, as observed in the outskirts of low mass clusters
\citep{walker2012a}. 
The projected temperature profile to the north is shown in Figure \ref{fig:projtprof}. 
We compare the {\it Suzaku} results with those from {\it Chandra} (this work) and {\it XMM-Newton} \citep{belsole2004}. 
We note that the {\it Chandra} results shown in Figure 
\ref{fig:projtprof} do not include the systematic uncertainties, and are
shown here for a rough check on the {\it Suzaku} temperature estimates.
We find good agreement between measurements from each satellite. We
note that both {\it Suzaku} and {\it Chandra} observations cover the radial
range out to 0.9~$R_{200}$ ($\sim$ 14$^{\prime}$, see Section
\ref{sec:mass}), and the best-fit temperatures measured by {\it Suzaku} and
{\it Chandra} are in agreement at the 1$\sigma$ confidence level. However, since {\it Suzaku} has a lower background at large radii ($\sim$$R_{200}$) and more precise temperature measurements (i.e., smaller systematic+statistical uncertainties), we will use {\it Suzaku} temperature and density measurements hereafter.

\begin{figure*}
\begin{center}
	\includegraphics[width=7.3in]{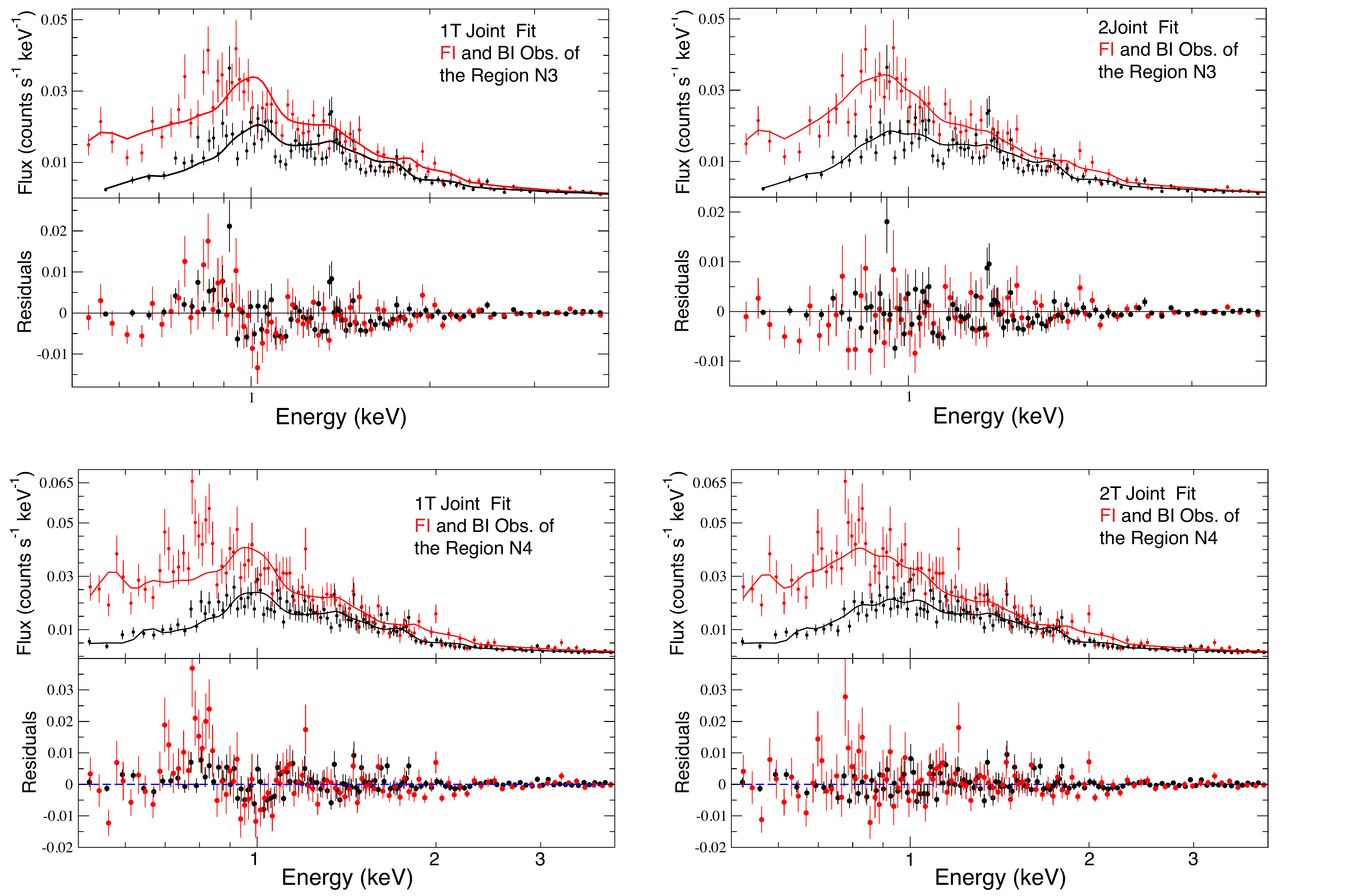}
	\caption{{\bf Left Panel:} Single temperature thermal model fit to the {\it Suzaku} joint FI (in red) and BI (in black) spectra extracted from the region N3 and N4 along the filament direction to the north (shown in Figure \ref{fig:image}). The best-fit model parameters obtained from this fit are given in Table \ref{table:1tfits}.  Residuals $<$2 keV indicate the probability of a secondary softer thermal component. {\bf Right Panel:} 2T \apec model fit to the joint {\it Suzaku} FI and BI spectra. The best-fit parameters of the fit are given in Table \ref{table:2tfits}. The change in $\Delta\chi^{2}$(N3,N4)=(28.5,24.4) for an extra two degrees-of-freedom suggests that the detection of a softer thermal component with $\sim$1 keV along the filament direction towards north is significant.}
	\label{fig:nn3}
\end{center}
\end{figure*}

The residuals in the spectrum after a model fit in the softer 
0.5 $-$ 2.0 keV band (shown in the left panel of 
Figure \ref{fig:nn3}) suggest the possible presence of a second, cooler thermal
component in the regions N3 and N4. To investigate this, we add another absorbed 
\apec component to the model (2T) and re-do the fit. Both the
temperature and the normalization of the second component are left
free, while the abundances are tied to each other between the two
\apec models. The best-fit parameters of the 2T \apec model and
the improvement in the fits are given in
Table \ref{table:2tfits}. Figure \ref{fig:nn3} (right panel) shows the improvements in the fits of both region N3 and N4. The temperature of the primary \apec
component increases from 1.98~$\pm$~0.09 keV to 2.93$_{-0.40}^{+0.57}$ keV, while the temperature of the secondary
component is estimated to be 0.99 $\pm$ 0.07 keV in region N3. The
change in the goodness of the fit statistics is significant, with a
$\Delta$C-Statistic of 64.5 for an additional two d.o.f. The C-Statistic value does not provide a statistical
test to quantify the significance of the improvement in the fit from adding the second component, thus we calculate the corresponding $\chi^{2}$ values before and after addition of the secondary \apec model. We find that adding two d.o.f. (additional temperature and its normalization) improves the $\chi^{2}$ by 28.5.
In region N4, the best-fit temperature of the primary \textit{apec} becomes 2.12$_{-0.37}^{+0.50}$ keV in the 2T \apec fits, while the temperature of the secondary component is 0.81~$\pm$~0.12 keV. The $\Delta\chi^{2}$ value of 24.4 with an additional two d.o.f., corresponding to a null hypothesis probability of $\sim10^{-6}$, suggests that the detection is significant. The best-fit parameters of these 2T models are summarized in Table \ref{table:2tfits}. The derived {\it XSPEC} normalizations, i.e. emission measures, and temperatures depend on the assumed metallicity. We provide the measurements of these observables for various solar abundance fractions. We note that the assumed metallically does not have a significant impact on temperature or emission measure of the hotter component in our fits. The discussion of the nature of this gas is provided in Section \ref{sec:fila}.

Considering that the calibration of XIS below 0.7 keV is uncertain, we re-perform the 1T and 2T model fits in the N3 and N4 regions to investigate the effect of this uncertainty on the temperature and normalization (i.e. density).  Fixing the abundance at 0.2 $A_\odot$, we find that the temperatures and normalizations of both models are consistent with results from the 0.5--7 keV band fits within the total (statistical plus systematic) uncertainties. The results from the 1T and 2T model fits in the 0.7--7 keV band are given in Table \ref{table:fits-0p7}. We conclude that the detection of the cooler $\sim$1 keV gas is not significantly affected by the effective area uncertainties below 0.7 keV.

\begin{table}
\def\arraystretch{1.7}
\begin{center}
\caption{The best-fit parameters of the 1T and 2T models in regions N3 and N4 in the 0.7-7 keV energy band}
\begin{tabular}{lccccc}
\hline\hline\\
Region	& $kT_{1}$	&$ \mathcal{N}_{1}$		& $kT_{2}$ 		&$ \mathcal{N}_{2}$	\\
	 	& (keV) 		&($10^{-6}$ cm$^{-5}$)	& (keV) 	& ($10^{-6}$ cm$^{-6}$) \\
	 \\
	 \hline
	 \\
N$_{3}$	& 1.96$\pm$0.18		& 16.45$^{+0.72}_{-0.74}$	& $-$ & $-$ \\
N$_{3}$	& 2.90$_{-0.66}^{+0.85}$	& 10.82$\pm$2.19	& 0.99$_{-0.10}^{+0.08}$	& 4.75 $\pm$ 1.21	\\
N$_{3}$	& 1.59$\pm$0.29		& 5.73$\pm$1.12 	& $-$ & $-$ \\
N$_{4}$	& 2.09$_{-0.51}^{+0.69}$ 	& 4.22$_{-1.16}^{+1.13}$	& 0.79$_{-0.13}^{+0.17}$ & 1.73$\pm0.44$ \\
\\
\hline\hline
\end{tabular}
\label{table:fits-0p7}
\end{center}
\end{table}

To investigate the X-ray emission along the filament to the south, we extract spectra
from two annular sectors (regions S1 and S2) extending south from the
center of A1750S. These regions are shown in Figure
\ref{fig:image}. Region S1 extends from the cluster core to
4$^\prime$, and region S2 extends from 4$^\prime$ to
9.7$^\prime$. The source counts in the combined FI and BI observations
are 2600 and 1700 in region S1, and
2200 and 2000 in region S2.  We first fit
the spectra with a 1T \apec model. The best-fit temperatures of 2.61~$\pm$~0.21 keV and 2.04$_{-0.41}^{+0.30}$ keV, and abundances of 0.19~$\pm$~0.08 A$_{\odot}$ and 0.20$_{-0.14}^{+0.24}$
A$_{\odot}$ are measured in regions S1 and S2, respectively. The results are shown in Table
\ref{table:1tfits} with the goodness of the fits. Abundance measurements of 0.2A$_{\odot}$ are consistent with the abundances measured in low mass systems \citep{walker2012}. The possible presence of the cool $\sim$ 1 keV gas is tested by performing 2T \apec fits. The additional secondary \apec model does not significantly improve the fits. Unlike the detection in the north, we find no evidence for such a component in the south.


\subsection{Off-Filament Direction}
\label{sec:offfilament}

The X-ray emission to the southeast, perpendicular to the putative large-scale filament, is examined using spectra extracted in annular sectors (SE1, SE2, and SE3 shown in Figure \ref{fig:image}) with radii of 0$^\prime$$-$4$^\prime$, 4.0$^\prime$$-$8.0$^\prime$, and 8$^{\prime}$ extending out to $R_{200}$ ($\sim$16$^\prime$) of the central sub-cluster. The total source counts in the FI and BI observations are 7800 and 5000 in region SE1, and 4000 and 2600 in region SE2, and 3000 and 2500 in region SE3.

To study the nature of the gas along the off-filament direction we
followed a similar approach to that outlined in
Section \ref{sec:fila}. The FI and BI spectra of each region are
first fit with a 1T \apec model. The best-fit parameters and the goodness
of these fits are given in Table \ref{table:1tfits}. The temperature
and abundance in the innermost region are 4.72$^{+0.17}_{-0.13}$ keV and 0.31~$\pm$~0.01 A$_{\odot}$. The best-fit temperature of the SE2 region is
4.83~$\pm$~0.40 keV. Unlike in region SE1, we are not able to constrain the abundance in region SE2, thus the abundance parameter is fixed at 0.2A$_{\odot}$. To test if the best-fit temperature is sensitive to the assumed metallicity, we perform the fit with abundances of 0.1A$_{\odot}$ and 0.3A$_{\odot}$.The best-fit temperature declines to 4.73~$\pm$~0.39 keV for an assumed abundance of 0.1A$_{\odot}$, while it increases to 4.95~$\pm$~0.38 keV for an abundance of 0.3A$_{\odot}$. However, the change in the measured temperature is not statistically significant.

\begin{figure*}[ht!]
\begin{center}
	\includegraphics[width=6.8in]{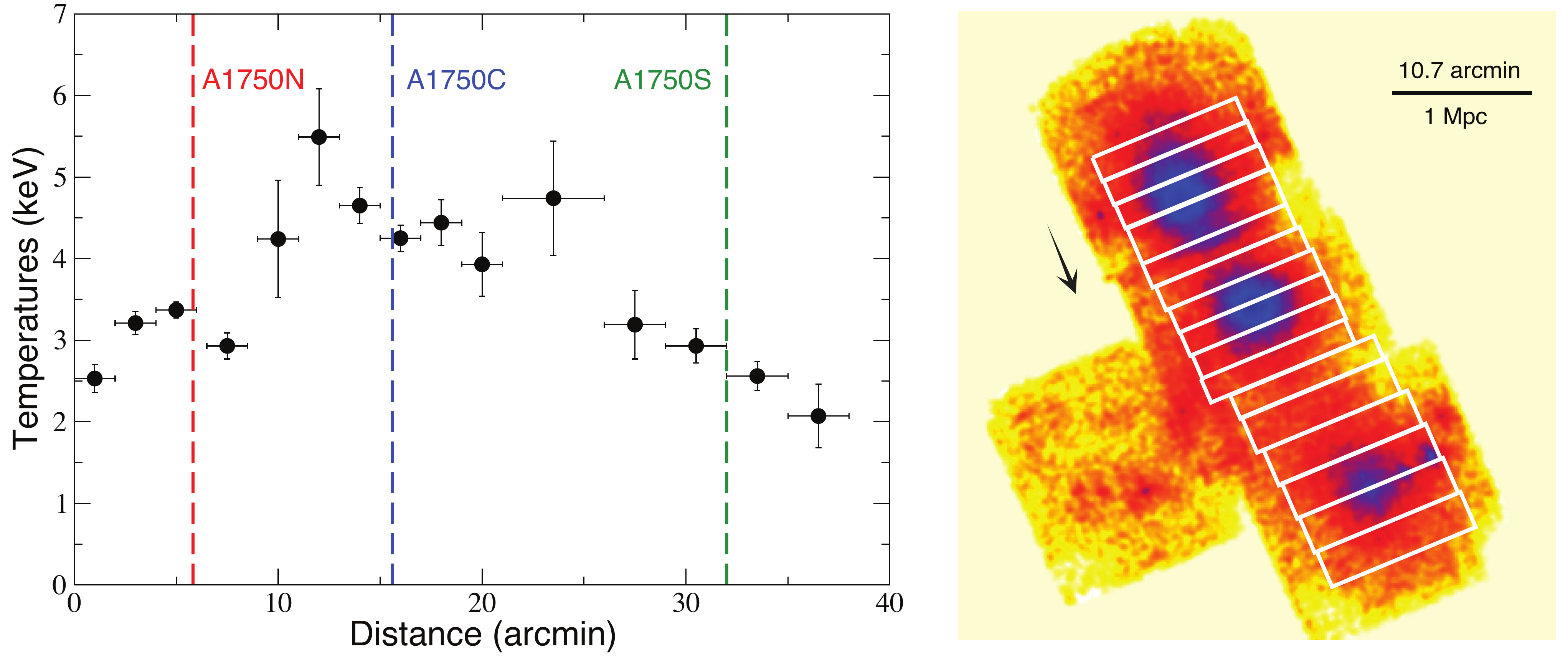}
	\caption{{\bf Left Panel:} Projected temperature measured from {\it Suzaku} observations as a function of radial distance along the filament axis. The dashed lines indicate the centroids of the sub-clusters A1750N, A1750C, and A1750S. {\bf Right Panel:} An image of the mosaicked {\it Suzaku} observations of the A1750 merger system. The spectral extraction regions are indicated in white. The selection of the regions is defined along the line connecting the centroids of the three sub-clusters. The direction is indicated with an arrow. }
	\label{fig:bridge}
\end{center}
\end{figure*}

The spectrum from region SE3 are dominated by the NXB above
5~keV. Therefore, we perform our fits in the 0.5$-$5 keV energy band
in this region. The best-fit temperature is 2.47$_{-0.68}^{+0.75}$ keV
for an assumed abundance of 0.2A$_{\odot}$. The temperature is
2.56$_{-0.70}^{+0.67}$ keV and 2.85$_{-0.74}^{+0.78}$ keV for fixed
abundances of 0.1A$_{\odot}$ and 0.3A$_{\odot}$, respectively. The
temperatures for our assumed abundances are all consistent within the
1$\sigma$ level. In all cases, we observe a significant sharp decline in the projected temperature at $\sim\ R_{500}$ (10.6$^{\prime}$; see Section \ref{sec:mass}) to the southeast.

Taking a similar approach as in Section \ref{sec:fila}, we fit
the spectra of the outermost regions SE2 and SE3 with a 2T \apec model. 
The temperature of the secondary component is
not constrained, and this addition
does not improve the fit significantly. Thus, we find
no evidence for a softer thermal component in the off-filament
direction. To further test if the $\sim$1 keV gas detected along the
filament to the north is observable along the off-filament
southeast direction, we scale the normalization of the
softer component detected in region N4 (see Table \ref{table:2tfits}) by the ratio of the
area of regions SE3 and N4. Freezing the normalization to the scaled
value of 1.5$\times\, 10^{-5}$ cm$^{-5}$ and the observed temperature
to 0.99 keV, we refit the FI and BI spectra of the SE3 region. The temperature and
normalizations of the primary component are unconstrained after the
fit is performed. The sharp decline in the goodness-of the fit
(C-Statistics value of 4931.15 for 471 d.o.f.) suggests that if the
$\sim$1 keV gas detected along the filament direction existed in this
region with the same surface brightness, it would be detected. Thus, this component is clearly absent in the off-filament direction.


\subsection{ICM Between the Sub-Clusters}
\label{sec:bridge}

We investigate the distribution of the gas temperature between
A1750N $-$ A1750C and between A1750C $-$ A1750S along the merger axis. We
define
rectangular regions along the line connecting the
centroids of the three sub-clusters (Figure~\ref{fig:bridge}, right), which are marked with dashed
lines in Figure~\ref{fig:bridge}, left.
We fit the spectra of the selected regions using a 1T
\apec model. Figure
\ref{fig:bridge} (left panel) displays the projected temperature as a function of
distance. We find that, starting from the northernmost region,
the temperature keeps rising towards the center of A1750N, and reaches a
peak temperature of 3.37 $\pm$ 0.10 keV. Due to the large PSF of
{\it Suzaku}, we cannot rule out or confirm the suggestion that A1750N is a
cool core cluster \citep{donnelly2001,belsole2004}. Continuing past A1750N,
the temperature rises up to 5.49 $\pm$ 0.59 keV with a sharp increase at $\sim$6$^\prime$
($\sim$ 0.5 Mpc). This increase in the temperature is significant at a level of 2.7$\sigma$. Hot, presumably shock-heated gas between A1750N and
A1750C, coinciding with the location where we detect hot gas with
{\it Suzaku}, has previously been observed in {\it Chandra} 
and {\it XMM-Newton} data \citep{molnar2013,belsole2004}. The presence of hot gas in
this region is an indication of an interaction between the A1750N and A1750C sub-clusters.

\begin{figure*}
\centering
\hspace{-4mm}\includegraphics[width=18.2cm, angle=0]{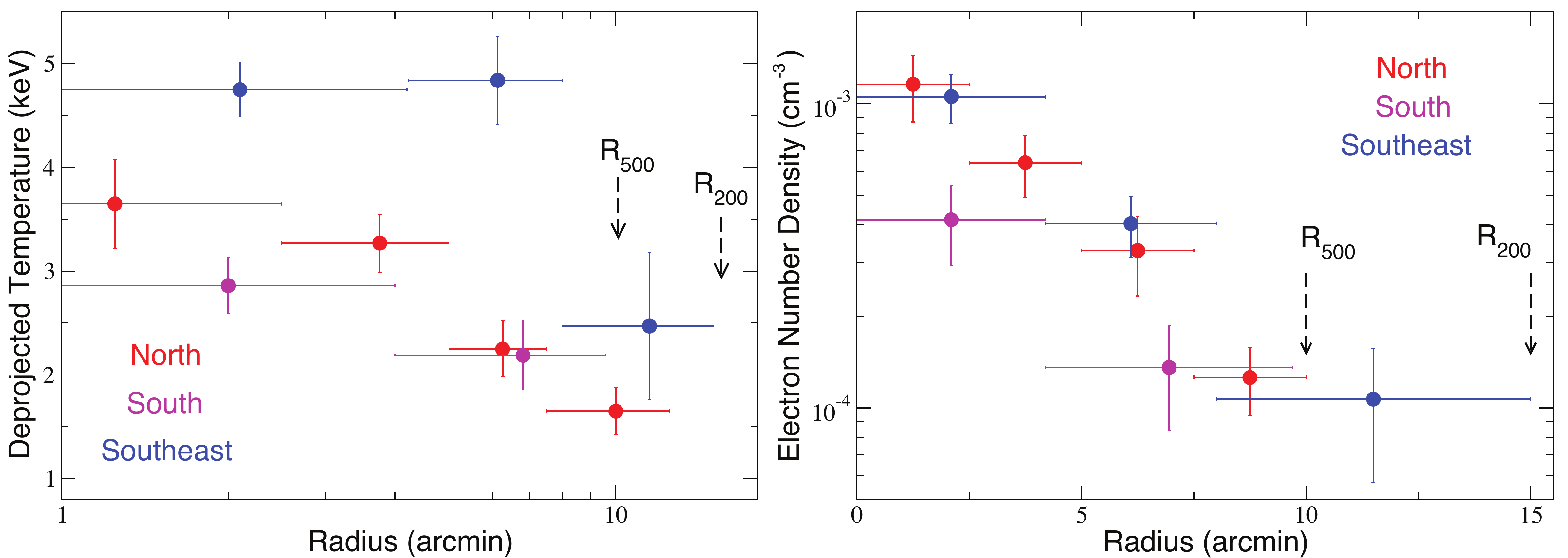}
\caption{Deprojected temperature (left panel) and electron density (right panel) profiles of the A1750 merger system to the north, south, and southeast directions. The 90\% systematic errors were added to the total error budget. The $R_{500}$ and $R_{200}$ estimates of A1750C are indicated with arrows.}
\label{fig:deprojprof}
\vspace{3mm}
\end{figure*}

A1750C shows a relatively uniform temperature around the centroid,
with a peak  temperature of 4.25 $\pm$ 0.16 keV. We detected another 
temperature peak located 7$^\prime$ away from A1750C, 
in the southwest direction, with a temperature of 4.74~$\pm$~0.70 keV. 
Southwest of this peak, the temperature declines to 3.19 $\pm$0.42 keV. This sharp 
decrease is significant at a 4$\sigma$ level, suggesting an interaction between the sub-clusters A1750S and A1750C.
A hot region, where the peak detected by {\it Suzaku} observations, was previously detected in the vicinity of A1750C \citep{belsole2004}. 
Due to large error bars on the temperature (5.7$^{+1.9}_{-1.7}$ keV), the authors were unable to determine the true nature of the structure and claimed that it could due to a point source.
Similarly a hot region was observed in {\it Chandra} data \citep{molnar2013} coinciding with the reported location of the peak. Here we confirm the extended nature of the emission and suggest a potential interaction between A1750C and A1750S. Although, we note that the optical data don't show any evidence of interaction between these clusters (see Section \ref{sec:optspec} for discussion).

The projected temperature continues to decline towards the center of the southern
sub-cluster A1750S. The central temperature of A1750S is 2.93~$\pm$~0.21
keV. The radial temperature profile shows that the temperature
decreases smoothly moving across the center of A1750S towards the southwest.

\section{Deprojected ICM Properties}
\label{sec:thermo}

To examine the radial profiles of  cluster masses and thermodynamical quantities such as
entropy and pressure, we determine the deprojected density and
temperature. The electron density is obtained from the best-fit
normalization $\mathcal{N}$ of the \apec model in \textit{XSPEC} using the relation,

\begin{equation}
\mathcal{N}= \frac{10^{-14}}{4\pi \, {D_{A}}^{2}\,(1+z)^{2}} \int n_{e}(r)\, n_{H}(r)\, dV \ \ \rm{cm^{-5}},
\end{equation} 
where $D_{A}$ is the angular size distance to the source in units of
cm, and $n_{e}$ and $n_{H}$ are the electron and hydrogen number
densities in units of cm$^{-3}$. We note that the ARFs generated by
\textit{xissimarfgen} assume a uniform source occupying an area of
400$\pi$ square arcminutes. We therefore apply a correction factor to each region and normalization prior to deprojection. An `onion-peeling' method is used to deproject the temperature and density profiles \citep{kriss1983, blanton2001, russell2008}. The resulting deprojected density and temperature profiles to the north, southeast, and south directions are shown in Figure \ref{fig:deprojprof}. 

We
extend the temperature and density profiles out to
0.9$R_{200}$ for A1750N to the north and $R_{200}$ for A1750C to the southeast with the new {\it Suzaku}
observations (see in Figure \ref{fig:image}). The temperature profiles to the north and southeast decline with radius and reach half of the peak value at $R_{200}$. Similar temperature declines have been reported for other clusters \citep[e.g.,][]{bautz2009,hoshino2010}. We observe a rather gradual decline in temperature to the north and south. However, the profile to the southeast indicates a uniform temperature within 8$^\prime$ and falls relatively rapidly beyond $R_{2500}$.

 
\subsection{Mass Analysis}
\label{sec:mass}

Based on the average deprojected density and temperature, we estimated the 
mass of each sub-cluster within $R_{500}$ using the $T_{X} - M_{tot}$ 
scaling relation \citep[][V09 hereafter]{vikhlinin2009}. The spectra between 0.15 $-$ 1$R_{500}$ are 
extracted to determine the global properties for each cluster. To avoid flux contamination, adjacent sub-clusters 
were excluded.

 A1750N has a best-fit global temperature of 3.14$^{+0.08}_{-0.07}$
 keV, and an abundance of 0.15 $\pm$ 0.03 A$_{\odot}$. Our measurement is consistent with the temperature of 3.17 $\pm$ 0.1 keV reported in \citet{belsole2004}. The scaling relation predicts a total mass of 1.98 $\times\ 10^{14}\ M_{\odot}$ at $R_{500}$ (9.3$^{\prime}$).
The best-fit temperature of A1750C
is 4.15$^{+0.12}_{-0.07}$ keV, with an abundance of 0.21$^{+0.03}_{-0.02}$
A$_{\odot}$. The global temperature reported in
\citet{belsole2004} is slightly lower ($kT~ =~ 3.87~ \pm~0.10$
keV). Their extraction region excludes the hotter plasma between
A1750N and A1750C, which may account for the difference observed in temperature. The V09 scaling relation predicts a total mass of 3.03~$\times\ 10^{14}\ M_{\odot}$ enclosed within $R_{500} (10.6^{\prime}$).
The spectral fit to A1750S gives a best-fit temperature of 3.59$_{-0.17}^{+0.20}$~keV and an abundance of
0.20$_{-0.06}^{+0.07}$. The estimated total mass within $R_{500}$ (9.9$^{\prime}$) is 2.43~$\times\ 10^{14}\ M_{\odot}$.

{\renewcommand{\arraystretch}{1.5}
\begin{table}
\begin{center}
\caption{Best-fit Parameters of the B10 Model}
\begin{tabular}{lcccccc}
\hline\hline\\

		&	North	&	Southeast\\
		\\
		\hline
$n_{e0}$ ($\times 10^{-3}$ cm$^{-3}$)	& 1.78 $^{+0.86}_{-0.41}$ 	& 2.19 $^{+0.92}_{-0.40}$  \\
$T_{e0}$ (keV)			& 3.89 $\pm$ 0.22		& 5.49 $\pm$ 0.27 \\
$n$					& 4.49 $^{+1.38}_{-0.52}$	& 6.01 $^{+1.79}_{-0.62}$\\
$r_{s}$ (arcmin)		& 300$^{*}$ 			& 480$^{*}$  \\
$\beta$ 				& 2.0$^{*}$			& 2.0$^{*}$\\
$\chi^{2}$ (dof)			& 5.35 (5)				& 3.21 (3)\\
\hline\hline
\multicolumn{3}{l}{%
  \begin{minipage}{7.3cm}%
    \scriptsize Fixed parameters are indicated with $^{*}$ %
  \end{minipage}%
} 
\\
\hline
\end{tabular}
\label{table:b10params}
\end{center}
\end{table}
}
\begin{figure*}
\centering
\includegraphics[width=18cm, angle=0]{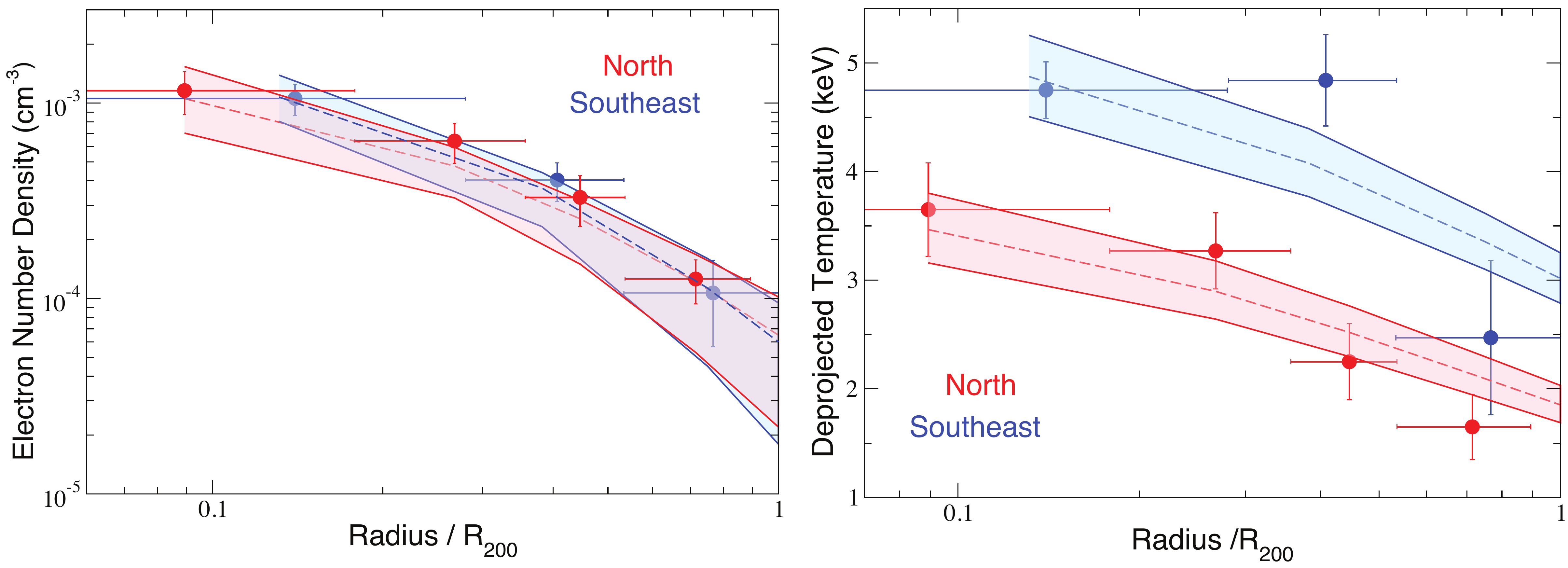}
\caption{B10 model fits to the density (left panel) and deprojected temperature (right panel) profiles to the filament (north) and off-filament (southeast) directions. The best-fit model is shown in dashed lines, while the 90\% confidence intervals are illustrated in shaded areas. The model provides an acceptable fits to the data, the goodness of the fits are given in Table \ref{table:b10params}.}
\vspace{3mm}
\label{fig:fits}
\end{figure*}

To investigate the radial behavior of the gas mass, the total mass,
and the gas mass fraction, we employ a physically motivated ICM model
described in \citet[][B10, hereafter]{bulbul2010, bulbul2011}. The B10 model is
based on the assumption that the ICM is a polytropic gas in
hydrostatic equilibrium in the cluster's gravitational potential. The
deprojected density and temperature profiles are fit simultaneously
using the B10 model. The fitting was performed using a Markov Chain Monte-Carlo (MCMC) approach, with Metropolis-Hastings sampling, to determine posterior distributions for the best-fit model parameters. The temperature profile is

\begin{equation}
T(r)=T_{0}\left[\frac{1}{(\beta-2)}\frac{(1+r/r_{s})^{\beta-2}-1}{r/r_{s}(1+r/r_{s})^{\beta-2}} \right],
\label{eqn_polytropic_temperature}
\end{equation} 

\noindent where the normalization constant $T_{0}$ is 

\begin{equation}
T_{0}=\frac{4\pi G \mu m_{p}}{k(n+1)}\frac{r_{s}^{2}\rho_{i}}{(\beta-1)}.
\label{eqn_relationof_T_n}
\end{equation}

Using the relation between temperature and gas density provided by the polytropic relation, 
the gas density is

\begin{equation}
n_{e}(r)= n_{e0} 
\left[\frac{T(r)}{T_{0}}\right]^{n},
\end{equation}

\noindent where $\beta$+1 is the slope of the total density
distribution, $n$ is the polytropic index, $r_{s}$ is the scale
radius, and $T_{0}$ and $n_{e0}$ are the central temperature and
density of the polytropic function. This model has sufficient fitting
flexibility to describe X-ray data, while making simple physical
assumptions
\citep{bonamente2012,hasler2012,bulbul2012,landry2013}. We note that
the core taper function in the B10 model is omitted in the fits
performed in this work, since {\it Suzaku} observations are not able to
resolve the cluster cores. Figure \ref{fig:fits} shows the best-fit models to the density (left panel) and temperature (right panel) in the off-filament and filament directions.

\begin{table*}
\centering
\caption{Gas and Total Mass Estimates at $R_{500}$ and $R_{200}$ Obtained from B10 Model}
\begin{tabular}{@{\extracolsep{\fill}}lcccccccc}
\hline\hline
Cluster          & $R_{500}$&$M_{gas}(R_{500})$    	& $M_{tot}(R_{500})$  & $f_{gas}(R_{500})$& $R_{200}$   &$M_{gas}(R_{200})$ & $M_{tot}(R_{200})$ & $f_{gas}(R_{200})$\\
                & (arcmin) & ($10^{13} M_{\odot}$) 	& ($10^{14} M_{\odot}$)&  & (arcmin)	& ($10^{13} M_{\odot}$) & ($10^{14} M_{\odot}$)  \\
\hline
\\
A1750N      & 9.3$^{\prime}$	& 1.86 $\pm$ 0.38	& 1.54 $^{+0.29}_{-0.26}$ &	0.12 $^{+0.04}_{-0.03}$ & 14.1$^{\prime}$	&  3.41 $^{+0.97}_{-0.92}$   &  2.32 $^{+0.43}_{-0.39}$ &  0.15 $^{+0.07}_{-0.06}$\\
\\

A1750C     &10.6$^{\prime}$	& 3.15 $^{+0.61}_{-0.63}$  & 3.04 $^{+0.56}_{-0.47}$&  0.10 $^{+0.04}_{-0.03}$& 16.2$^{\prime}$&  5.46 $\pm$ 0.16 & 4.85 $^{+1.62}_{-1.18}$ & 0.11 $^{+0.10}_{-0.06}$ \\
\\
\hline  \hline
\label{table:mass}
\end{tabular}
\end{table*}

Due to the limited number of data points compared to the number of free
model parameters of the B10 model (five in this case), we were not
able to constrain all of the free parameters of the model. The $\beta$
parameter is fixed to the slope of the Navarro-Frenk-White profile
\citep{navarro1996}, while the scale radius r$_s$ (fixed in our fits),
the radius beyond which the temperature starts declining, is
estimated from the temperature profiles (see Figure
\ref{fig:fits}). The rest of the model parameters ($n$, $n_{e0}$, and
$T_{e0}$) are allowed to vary independently. The best-fit parameters
of the model are given in Table \ref{table:b10params}, along with the
goodness of the fits. The best-fit models for the density and
temperature profiles are displayed in Figure \ref{fig:fits}, with 90\%
confidence intervals. Given the limited number of data points, the profiles to the south are not constrained.

The total mass enclosed within radius $r$ is

\begin{equation}
\begin{aligned}
M(r)=\frac{4\pi\rho_{i}r_{s}^{3}}{(\beta-2)}\left[ \frac{1}{\beta-1} +\frac{1/(1-\beta) - r/r_s}{(1+r/r_{s})^{\beta-1}}\right].
\end{aligned}
\label{eqn:totalmass}
\end{equation}

\noindent The normalization factor for the total matter density is $\rho_{i} = [T_{0} k (n + 1)(\beta- 1))/(4\pi G \mu m_{p}rs^{2}]$. 

The gas mass $M_{gas}$ is computed by integrating the gas density profile within the volume,

\begin{equation}
M_{gas}(r) =  4\pi \mu_{e} \ m_{p}\int n_{e}(r) \ r^{2} \ dr,
\label{eqn:gasmass}
\end{equation}

\noindent where $\mu_{e}$ and $m_{p}$ are the mean molecular weight per electron and the proton mass. 

The gas mass fraction is 

\begin{equation}
f_{gas} = \frac{M_{gas}}{M_{tot}}.
\end{equation}

\begin{figure*}
\centering
\includegraphics[width=18cm, angle=0]{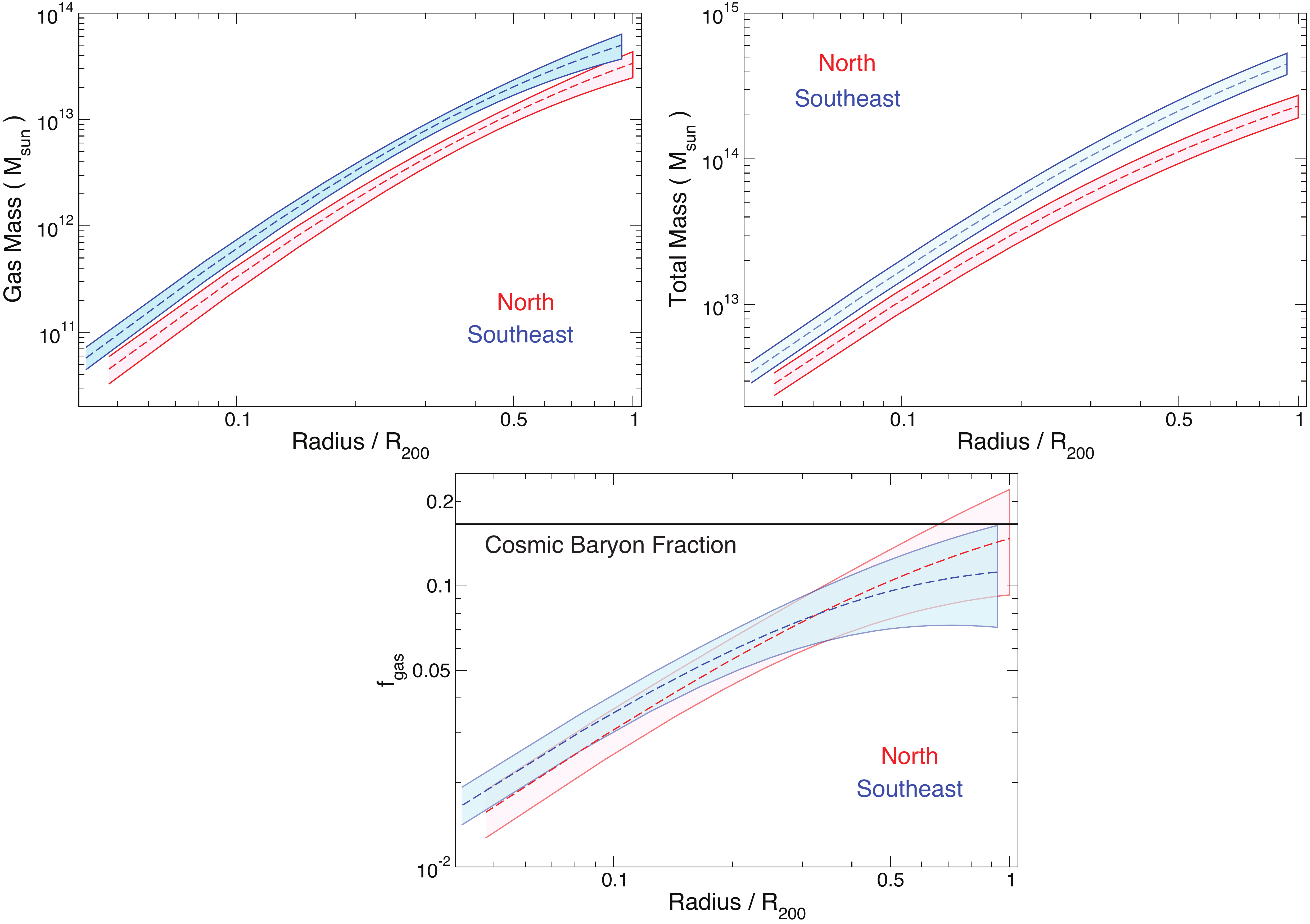}
\caption{Gas mass, total mass, and gas mass fraction obtained from the
  B10 model of A1750N and A1750C subclusters. The dashed lines
  show the masses obtained from the best-fit models. The shaded area
  shows the 90\% confidence interval. {\it Suzaku} data indicates that the
  gas mass fraction at $R_{200}$ is consistent with the cosmic baryon
  fraction of 0.166 indicated by a solid line in the lower panel
  \citep{komatsu2011}. We do not see any evidence for super-cosmic
  values for $f_{gas}$ that would arise from clumping at large radii either along or perpendicular to the large scale filamentary structure. }
\vspace{3mm}
\label{fig:mass}
\end{figure*}

The gas mass, total mass, and $f_{gas}$ are measured at
$R_{500}$, determined using the V09 scaling relations, and are given in
Table \ref{table:mass}. 
Following \citet{pratt2010}, we assume
$R_{500}$~=~0.659$R_{200}$. The total mass, gas mass, and gas mass
fraction profiles are plotted in Figure \ref{fig:mass}.
We find that the total masses enclosed within $R_{500}$ are well
within agreement with the total masses estimated using the V09 scaling
relations. The gas mass fractions of A1750N and A1750C are consistent
with the gas mass fraction expected for clusters in this mass range
based on the V09 scaling relations ($f_{gas}\,\sim$ 0.11) at $R_{500}$.  
The B10 model was then used to calculate the masses and mass fraction at $R_{200}$. We found that the gas mass fraction of A1750C and A1750N at $R_{200}$ is 0.11$^{+0.10}_{-0.06}$ and 0.15$^{+0.07}_{-0.06}$.

The virial masses of the A1750N and A1750C subclusters are in agreement with the mass estimates from the optical observations at a 2.7$\sigma$ level (see Section \ref{sec:optspec}). However, we note that, the cluster mass inferred from X-ray analysis depends on the geometry of the merger, hydrostatic equilibrium, and other model parameters (e.g. scale radius) of the merging clusters. 

The gas fractions derived in the filament and off-filament directions are consistent with 
the cosmic baryon fraction derived from WMAP seven-year data of 0.166 \citep{komatsu2011}. 
Similarly, gas mass fractions consistent with the cosmic value, were
observed in RX J1159+5531 \citep{humphrey2012,su2015},
A1689 \citep{kawaharada2010}, and A1246 \citep{sato2014}.
However, we note that the total mass estimates are based on a few
assumptions on the distribution of the gas properties. Spherical
symmetry and isotropy are assumed when calculating these masses. Such assumptions may bias our results, particularly
in a merger system at large radii. 

\subsection{Entropy Profiles}
\label{sec:entro}

\begin{figure*}
\centering
\hspace{-6mm}\includegraphics[width=18.6cm, angle=0]{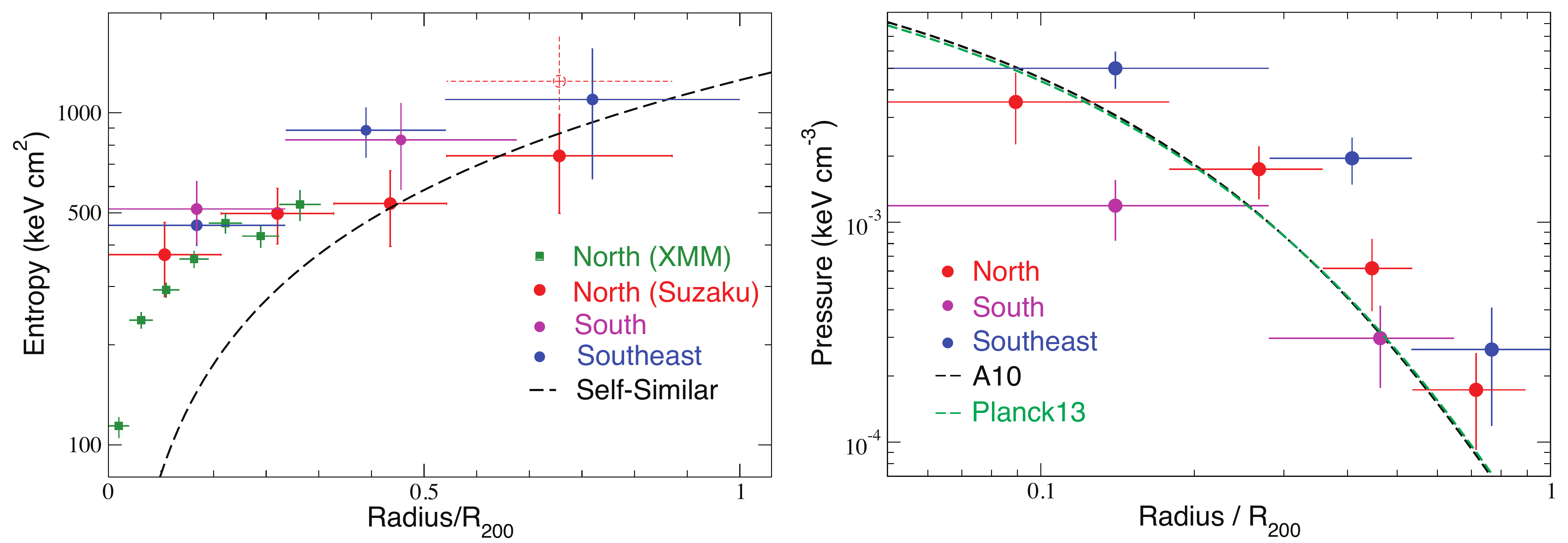}
\caption{{\bf Left Panel:} Entropy profile in the filament (to south and north), and off-filament (to southeast) directions. {\it XMM-Newton} observations are plotted in green squares to the north are in good agreement with the {\it Suzaku} observations \citep{belsole2004}. The dashed line indicates the self-similar expectation \citep{voit2005, pratt2010}. The entropy exceeds the self-similar model within the inner $\sim 0.5 R_{200}$ and follows the expectation beyond this radius. The entropy becomes more consistent when the contribution from the cool gas detected to the north is removed. The entropy of the hotter component is shown with a data point in red with dashed lines. {\bf Right Panel:} Pressure profiles in the filament, and off-filament directions. The Universal \citet{arnaud2010} and \citet{planck2013} profiles are shown in black and green dashed lines, respectively. Both profiles are scaled to the estimated $R_{200}$ of each sub-cluster.}
\vspace{3mm}
\label{fig:deprojentropress}
\end{figure*}

The entropy ($K = kT/n_{e}^{2/3}$) and pressure ($P =
n_{e} kT$) profiles are calculated using the electron
density ($n_{e}$) and deprojected temperature ($kT$). The profiles along the
filament and off-filament directions are shown in Figure \ref{fig:deprojentropress}.

In the absence of non-gravitational processes, such as radiative cooling and
feedback, cluster entropy profiles are expected to follow the simple power-law relation 

\begin{equation}
\frac{K}{K_{500}} = 1.42 \ (R/R_{500})^{1.1},
\end{equation}

\noindent where we assume a cosmic baryon fraction of $f_b$ = 0.15, with a characteristic entropy of
\begin{equation}
\begin{split}
K_{500} = 106\ \rm {keV \ cm^{-2}} \left( \frac{M_{500}}{10^{14}\ h_{70}^{-1}\ M_{\odot}}\right)^{2/3}\left(\frac{1}{f_{b}}\right)^{2/3} \\
\times \ E(z)^{-2/3}  \ h_{70}^{-4/3} \ \ \rm {keV \ cm^{-2}},
\end{split}
\end{equation}

\noindent  \citep{voit2005, pratt2010}. We used an $M_{500}$ (the
total mass within $R_{500}$ of A1750C) of 3 $\times\ 10^{14}\ M_{\odot}$, as
determined in \S\ref{sec:mass}. The resulting expected self-similar
entropy profile for A1750C is shown as the dashed lines in Figure
\ref{fig:deprojentropress} (left).
 
We find that the entropy along the filament directions (to the north and south)
and off-filament direction derived from {\it Suzaku} data alone are in good
agreement with each other within $< 0.3R_{200}$. Profiles obtained from {\it XMM-Newton} observations 
are consistent with those from {\it Suzaku} data within $<0.2R_{200}$. The observed entropy
exceeds the self-similar model prediction within $<0.5R_{200}$, which
we attribute to the influence of non-gravitational processes (e.g., AGN feedback, infalling substructures due to violent merging events) in the
subcluster cores. Such an influence on the 
entropy profiles of a sample of low redshift clusters ($z <$ 0.25) was
reported by \citet{walker2012c}.

The entropy profiles follow a 
flatter profile beyond a radius of $\sim 0.3\ R_{200}$, and become 
consistent with the self-similar model, both along the northern filament
and off-filament directions. We find that the entropy profile towards the northern
filament reaches the self-similar level at smaller radii ($\sim$0.4R$_{200}$) as compared
with the off-filament direction. This may be due to the lower temperature 
gas ($\sim$1 keV) observed to the north, which biases the average temperature low, and depresses the measured value of the entropy.
The entropy profile along the off-filament direction stays above the
self-similar expectation to $\sim 0.5\ R_{200}$. Beyond this radius it remains 
consistent with the self-similar prediction. If the entropy contribution from the cool gas detected to 
the north is removed, the entropy rises to 1245.6 $\pm$ 486.5 keV cm$^2$
(shown in Figure \ref{fig:deprojentropress} with the dashed data point in red) and becomes more consistent with the entropy to the southeast. 
This provides evidence that the cool gas does indeed lead to a slight decrease in the entropy, although not at the level seen in other systems where it is likely arises from gas clumping \citep{urban2014}.

Unlike the rising, self-similar entropy observed in A1750, a flattening
of entropy profiles near $R_{200}$ appears to be a common feature in
other relaxed and disturbed clusters \citep[for a review, see][]{reiprich2013}. A few detailed studies of nearby bright merging systems have probed the physical properties of the ICM at large radii, e.g. the Coma cluster \citep{s2013} and the Virgo cluster \citep{urban2011}. {\it XMM-Newton} observations of the dynamically young cool Virgo cluster revealed a suppressed entropy profile beyond 450 kpc by a factor of 2--2.5 below the expectation from pure gravitational collapse models. Authors attributed this flattening to gas clumping at large radii. In the merging Coma cluster, \citet{s2013} find no evidence for entropy flattening along the relatively relaxed directions, although due to large uncertainties they are unable to exclude entropy flattening at the level of what is observed in some relaxed clusters. 

There has been great effort in the literature to explain the seeming ubiquity of 
flattened entropy profiles at large radii. In the
hierarchical model of structure formation, clusters form by accreting
material from their surrounding large-scale structure. Accretion of
infalling subhalos can cause gas motions and ``clumpiness" around $R_{200}$. 
These subhalos tend to have lower temperature and higher density than the
surrounding ICM, leading to a bias towards lower temperatures and higher densities in the
emission measure-weighted spectra, if the subhalos are unresolved. 
The level of gas inhomogeneities is characterized through the clumping factor 
($C = <n_{e}^2>/ <n_{e}>^{2}$). As a result of overestimation of density, the gas mass, and subsequently the gas mass fraction, are biased high (i.e. above the cosmic baryon fraction).
The observed excess in the gas mass fraction ($M_{gas}/ M_{tot}$) in the {\it Suzaku} observations 
of the Perseus cluster was explained with a very large clumping factor of 3 $-$ 4 around $R_{200}$ \citep{s2011}. 

\citet{nagai2011} reported that the expected clumpiness factor at $R_{200}$ can be as large as 2 and confirmed 
the flattened entropy profiles beyond $r\ >\ 0.5R_{200}$ in their non-radiative and cooling+star-formation simulations.
However, \citet{walker2012c} examined entropy profiles for a sample of relaxed clusters at $z\ <$ 0.25 out to $R_{200}$ and concluded that the gas clumping calculated in the numerical simulations is insufficient to reproduce the observed flattening of the entropy.

An alternative explanation to the flattening was proposed by
\citet{hoshino2010} and \citet{akamatsu2011}, and is based on the
electron-ion non-equilibrium in the cluster outskirts.
 If the energy is not transferred to the electrons
through electron-ion collisions sufficiently rapidly, the electron
temperature remains low compared to that of ions, leading to an
apparent entropy suppression at $R_{200}$.

\citet{lapi2010} and \citet{cavaliere2011a} proposed that the
flattening in the entropy is a result of a weakened accretion shock as
it expands. The bulk energy carried along with the shock increases the
turbulence and non-thermal pressure support in the outskirts, but the
shock is not energetic enough to raise the intra-cluster entropy. The decreasing thermalization in low-density regions results in a tapered entropy around $\sim$$R_{200}$. This claim supports the observed azimuthal variations in entropy in cool-core clusters \citep{ichikawa2013} and in the non-cool core Coma cluster \citep{s2013}. 
Other proposed explanations of entropy flattening include a rapid radial fall of the gas temperature caused by non-gravitational effects \citep{f2014} and cosmic-rays consuming as a significant sink for the kinetic energy in the outskirts \citep{fujita2013}.

On the other hand, \citet{eckert2013a} have performed a joint Planck SZ and ROSAT X-ray analysis of 18 galaxy clusters and concluded that entropy profiles are consistent with a self-similar power-law increase expected from pure gravitational infall. The discrepancy between the \citet{eckert2013a} and the \citet{walker2012c} results is due to the differing dependence on SZ and X-ray signals to the electron pressure used to derive entropy profiles \citep{f2014}. 

Self-similar entropy profiles at $R_{200}$ have been previously observed in low mass relaxed fossil groups, e.g. RX J1159+5531 \citep{humphrey2012,su2015}. On the other hand, the entropy of morphologically relaxed groups has been found to be significantly higher than self-similar at $r< R_{500}$ \citep{sun2009}.
However, massive mergers ($M_{200}>10^{14}\ h^{-1}M_{\odot}$) are expected to
have a higher level of gas clumping, since they have a larger fraction
of lower-temperature gas that is not detectable in the X-ray band
\citep{nagai2011,vazza2013}. Although A1750 is a dynamically young,
massive system, we do not find evidence for gas clumping in this merger system. Entropy profile measurements along the off-filament and filament directions are in agreement with each other and with the universal expectation with a power-law relation $\propto \, r^{1.1}$. Remarkably in A1750, the entropy 
profiles within $R_{200}$ do not seem to have been influenced by the apparent filamentary structure of the system. Our results suggest that gravitational collapse is the main driver of the temperature and density profiles in the outskirts.

\subsection{Pressure Profiles} 
\label{sec:press}

We also examine the pressure profiles along the off-filament and
filament directions. Pressure profiles are calculated assuming an
ideal gas law with $P(r)=n_{e}(r)\, k\, T(r)$, and compared to the universal
pressure profiles of \citet{arnaud2010} (A10, hereafter) and \citet{planck2013} (Planck13, hereafter) for clusters with mean redshifts of 0.11 and 0.17, respectively. The A10 universal pressure profile is

\begin{equation}
\begin{aligned}
P(r)= &\, 1.65 \times 10^{-3} \, h(z)^{8/3}\left[ \frac{M_{500}}{3\times10^{14}\,h_{70}^{-1}\,M_{\odot}}\right]^{2/3+\kappa}\\
&\times \mathbb{P}(r/R_{500})\ h_{70}^{2} \  \ \rm{keV}\ {cm^{-3}},
\end{aligned}
\label{eqn:arnaud}
\end{equation}
 
\noindent where the scaled pressure profile is characterized based on the generalized Navarro-Frenk-White profile \citep{nagai2007} 

 \begin{equation}
\mathbb{P}(r/R_{500}) = \frac{P_{0}}{(c_{500}\,r/R_{500})^{\gamma}[1+(c_{500}\,r/R_{500})^{\alpha}]^{(\beta-\gamma)/\alpha}},
\end{equation} 

 \noindent and

 \begin{equation}
 \begin{split}
 \kappa	&= \alpha_{p}+\alpha^{\prime}_{p}(r/R_{500})\\
 		&= (a_{p}+0.10)-(a_{p}+0.10)\frac{(r/0.5\, R_{500})^{3}}{1+(r/0.5\, R_{500})^{3}},
		\label{eqn:kappa}
\end{split}
 \end{equation}

\noindent with best-fit parameters of
$P_{0}$ = 8.403$\,{\rm h_{70}^{-3/2}}$, $c_{500}$ = 1.177, $\gamma$ =
0.3081, $\alpha$ = 1.0510, and $\beta$ = 5.4905. The first term in
Equation \ref{eqn:kappa}, $\alpha_{p}$, is an approximation which
depends on the departures from the standard scaling relations, while
the second term, $\alpha^{\prime}_{p}$, represents a break from
self-similarity. Since non-gravitational processes become less
dominant at large radii, the latter term is negligible at
$>R_{500}$. The A10 universal pressure profile primarily samples the
inner regions, while the Planck13 profile samples the cluster
outskirts. The pressure profiles derived from the Planck observations for a sample of 62 galaxy clusters found slightly higher
pressure than that predicted by A10 in the outskirts of
clusters. These profiles were obtained by averaging pressure profiles
from all azimuths for a large sample of clusters with different
dynamical states. The dispersion over the universal profiles can be as large as 100 \% at $\sim$$R_{500}$ (see Figure 8 in A10).

We compare the pressure profiles of A1750 with the universal profiles of A10 and Planck13
in the right panel of Figure \ref{fig:deprojentropress}. While the
pressure profile along the filament direction to the north agrees with
the universal profile, the profile along the
off-filament direction is higher and the profile to the south is lower than the expectation within $<$ 0.2$R_{200}$. On the other hand, the profile in the filament direction to the south and to the north
is consistent with the A10 and Planck13 universal profiles at large radii ($\sim$$R_{200}$) at the 2.7$\sigma$ level. The pressure to the southeast exceeds the universal models at all radii. 

Pressure excesses at large radii have
been previously reported in other relaxed clusters, e.g. PKS 0745-191
\citep{walker2012a}, the Centaurus cluster \citep{walker2013b}, and
the fossil group RX J1159+5531 \citep{humphrey2012}, and were
attributed to gas clumping. Figure \ref{fig:deprojprof} indicates that the excess in the pressure along the southeast direction compared to the north or south directions in A1750 is due to high temperature (not high density). On the contrary, clumping (if it existed in this system) would bias the density measurements high, leading to an excess in pressure and a decrement in entropy in the outskirts. Therefore, the deviation from the universal profile in A1750 is unlikely to be due to clumpy gas, since other evidence for clumping, e.g. entropy flattening
and an excess in gas mass fraction (see Section \ref{sec:mass}), are not observed in this system. 
We note that \citet{belsole2004} reported the detection of a weak $M=1.2$, shock resulting from a merger event intrinsic to A1750C along the southeast direction. This merger event may elevate the temperature and cause deviations from the universal profile.
In any event, given the large dispersion among pressure profiles of clusters in the A10 and Planck13 samples, we do not expect the pressure profiles derived in A1750 in perfect agreement with their results.

\subsection{Nature of the Cool Gas Detected to the North}
\label{sec:coolgas}

The cool $\sim$ 1 keV gas detected in regions N$_{3}$ and N$_{4}$ (see Figure \ref{fig:nn3}) may be 1) the hot dense WHIM connecting A1750N to the large-scale filament, 2) stripped ICM gas
formed as a result of infalling groups, or 3) gas stripped from A1750N itself, as it interacts with filament gas or with A1750C. The feature is relatively
extended with an observed radial range of $>$ 0.62 Mpc. 
Assuming a geometry for the merger system, the mass of the 
feature can be calculated (see Section \ref{sec:dynamics} for the
detailed calculation). Assuming that the density of the feature is
constant within each region (5.56 $\times\, 10^{-6}\, ({\it l}/1\,
\rm{Mpc})^{-0.5}$ cm$^{-3}$), and can be described as a cylinder that extends to 1.2 Mpc 
with a line-of-sight depth of the structure ({\it l}), we obtain a gas mass of 4.13 $\times\, 10^{11}\,  ({\it l}/1\, \rm{Mpc})^{0.5}~ M_{\odot}$.
The observed flux of the feature (1 $\times\, 10^{-15}$ erg cm$^{-2}$ s$^{-1}$ arcmin$^{-2}$), density, and temperature ($\sim$0.8 keV) are consistent with the expected surface brightness and temperature of the dense portion of the WHIM, where the large-scale structure interacts with the cluster's ICM \citep{dolag2006,werner2008, planck2013fila}. 

Such a filamentary feature also may be due to an additional small subcluster infalling 
into A1750N which is being disrupted as it interacts with the main cluster. The bulk of
the halo gas lags behind the infalling groups, and is stripped by the ram
pressure of the ambient ICM. Such halos are expected to have an
average temperature of $\sim$1 keV with a typical halo mass of
3$\times\, 10^{13}~ M_{\odot}$ \citep{sun2009}. Bright, large-scale
($\sim$ 700 kpc) stripped tails have been observed in the outskirts of
galaxy clusters, e.g. the Virgo cluster \citep{randall2008}, A85
\citep{ichinohe2015}, and A2142 \citep{eckert2014}. The stripped gas
from infalling halos may seed gas inhomogeneities (i.e. clumping), which
suppress the average entropy inferred at large radii. In such systems,
a flattening of the entropy profile, as well as an excess in the gas
mass fraction as compared with the cosmic value, have often been
observed in cluster outskirts. In the case of A1750, the entropy profile remains consistent with the self-similar prediction out to $R_{200}$, and the gas mass fraction is consistent with the cosmic value (see Section \ref{sec:mass}), implying that the observed cool gas could indeed be the densest and hottest parts of the warm-hot intergalactic medium. In addition, dense and cool clumps in the outer cluster regions are expected to lead to more entropy flattening since they will lower the average temperature, and, more importantly raise the average density \citep{walker2013}. Along the north direction of A1750N, there is sufficient cool gas to be detected, but it does not cause the dramatic entropy flattening seen in some other clusters, suggesting that its density cannot be too high.
Completely ruling out the ram pressure-stripping scenario for this cool gas requires deeper {\it Chandra} observations with good angular resolution. The WHIM interpretation of this feature cannot be firmly established based on the {\it Suzaku} data.

\section{Summary}
\label{sec:conc}

We present an analysis of the strongly merging cluster A1750 using
{\it Suzaku} and {\it Chandra} X-ray observations, and {\it MMT} optical observations out to
the cluster's virial radius. The deep {\it Suzaku} observations allow us
to constrain the entropy, pressure, and mass profiles at the outskirts, both along and perpendicular to the large-scale filament. We use optical observations to constrain the dynamical state of the
cluster. Our major results are:

\begin{enumerate}
\item A1750N and A1750C have a  78\% chance of being bound. There is an apparent hot region with a temperature of 5.49 $\pm$ 0.59 keV in between these subclusters implying an interaction. The red galaxy distribution and the velocity dispersion data prefer a pre-merger scenario. In an early pre-merger scenario, one expects the outer ICM atmospheres of the subclusters to interact subsonically, driving shocks, and ultimately creating a heated ICM region between the subclusters, e.g., N7619 and N7626 \citep{randall2009}.

\item We find overall a good agreement between the measured entropy
  profiles and the self-similar expectation predicted by gravitational collapse near
  $R_{200}$ both along and perpendicular to the putative large-scale
  structure filament. Unlike some other clusters, the entropy profiles at large radii, both perpendicular and along the filamentary directions, are consistent with each other. Agreement of the entropy with the self-similar expectation at
  $R_{200}$ in this 
  massive and dynamically young system suggests that A1750
  exhibits little gas clumping at large radii.

\item The gas mass fractions in both the filament and off-filament directions are 
	consistent with the cosmic baryon fraction at $R_{200}$. This may indicate that gas clumping may be less common in such smaller, lower temperature ($kT~\sim$~4 keV) systems \citep[with a few exceptions, e.g. the Virgo cluster,][]{urban2011}. Cluster mass may therefore play a more important role in gas clumping than dynamical state.

\item An extended gas ($>$ 0.62 Mpc) is observed to the north of the A1750N subcluster along the large scale structure, where one would expect to detect the densest part of the WHIM in a filament, near a massive cluster. The measured temperature (0.8 $-$1 keV), density, and radial extent of this cool gas is consistent with the WHIM emission. The thermodynamical state of the gas at that radius (i.e. self-similar like entropy profile, and gas mass fraction consistent with the cosmic value) favors the WHIM emission interpretation. However, a deeper observation with {\it Chandra} resolution is required to distinguish this diffuse filamentary gas from an infalling substructure, or gas from ram pressure-stripping.
	
\end{enumerate}

\section{Acknowledgements}
We thank Gabriel Pratt for kindly providing temperature, density, and entropy profiles from {\it XMM-Newton} data. We also thank Mike McDonald and John Zuhone for useful comments and suggestions. EB was supported in part by NASA grants NNX13AE83G and NNX10AR29G.
SWR was supported by the {\it Chandra} X-ray Center
through NASA contract NAS8-03060 and by the Smithsonian Institution. MBB acknowledges support from the NSF through grant AST-1009012. ELB and RPM was partially supported by the National Science Foundation through grant AST-1309032. CLS was funded in part by {\it Chandra} grants GO4-15123X and GO5-16131X and NASA XMM grant NNX15AG26G. AEDM acknowledges partial support by {\it Chandra} grants GO2-13152X and GO3-14132X. Authors thank Prof. Dr. Nihal E. Ercan for providing the support for CE.

\end{document}